\documentclass[onecolumn,preprint,aps,prl,superscriptaddress,floatfix]{revtex4}
\usepackage{comment}
\usepackage[T1]{fontenc}
\usepackage[utf8]{inputenc}
\usepackage[english]{babel}
\usepackage{multibib}
\usepackage[a4paper,top=2cm,bottom=2cm,left=3cm,right=3cm,marginparwidth=1.75cm]{geometry}
\usepackage{color}
\usepackage{float}
\usepackage{amsmath}
\usepackage{amstext}
\usepackage{amssymb,bm}
\usepackage{graphicx}
\usepackage{bbold}
\usepackage{hyperref}

\newcommand{\be}{\begin{equation}gin{equation}}
\newcommand{\ee}{\end{equation}}

\newcommand{\ek}{\varepsilon_\mathbf{k}}

\def\pd#1#2{\frac{\partial #1}{\partial #2}}

\def\f0{f^0_\mathbf{k}}

\usepackage{soul}
\makeatletter
\date{}

\makeatother
\setul{0.5ex}{0.3ex}
\definecolor{Blue}{rgb}{0,0.0,1}
\setulcolor{Blue}
\usepackage{babel}

\begin{document} 
\author{Tatiana G. Rappoport}
\affiliation{University of Lisbon and Instituto de Telecomunicações, Avenida Rovisco Pais 1, Lisboa, 1049-001 Portugal}	
\affiliation{Instituto de F\'\i sica, Universidade Federal do Rio de Janeiro, C.P. 68528, 21941-972 Rio de Janeiro RJ, Brazil}
\author{Tiago A. Morgado} 
\affiliation{Instituto de Telecomunicações and Department of Electrical Engineering, University of Coimbra, 3030-290 Coimbra, Portugal} 
\author{Sylvain Lanneb\`ere}
\affiliation{Instituto de Telecomunicações and Department of Electrical Engineering, University of Coimbra, 3030-290 Coimbra, Portugal} 
\author{M\'ario G. Silveirinha}
\affiliation{University of Lisbon and Instituto de Telecomunicações, Avenida Rovisco Pais 1, Lisboa, 1049-001 Portugal}	

\title{Engineering transistor-like optical gain in two-dimensional materials with Berry curvature dipoles}
\begin{abstract}
Semiconductor transistors are essential elements of electronic circuits as they enable, for example, the isolation or amplification of voltage signals. While conventional transistors are point-type (lumped-element) devices, it may be highly interesting to realize a \emph{distributed} transistor-type optical response in a bulk material. Here, we show that low-symmetry two-dimensional  metallic systems may be the ideal solution to implement such a distributed-transistor response. To this end, using the semiclassical Boltzmann equation approach, we characterize the optical conductivity of a two-dimensional material  under a static electric bias.
It is found that similar to the nonlinear Hall effect, the electron transport depends on the Berry curvature dipole.  Our analysis reveals that the electro-optic effect modifies the optical conductivity of the material, breaking the electromagnetic reciprocity and yielding a dynamical response that imitates that of a transistor but in a distributed volume. Furthermore, the effective conductivity tensor can be non-Hermitian, opening the possibility of optical gain. To maximize the non-Hermitian response, we explore the specific case of strained twisted bilayer graphene. Our analysis reveals that the optical gain for incident light transmitted through the biased system depends on the light polarization, and can be quite large, especially for multilayer configurations.
  \end{abstract}
\maketitle
Lorentz's reciprocity is at the origin of the bidirectional nature of conventional photonic devices, limiting the ways that electromagnetic signals can be manipulated in such systems \cite{Pozar}. To bypass this limitation and produce a nonreciprocal electromagnetic response, it is necessary to apply a suitable bias to the system \cite{Caloz2018,Asadchy2020}. This can be achieved by breaking time-reversal (TR) symmetry \cite{Haldane2008}, traditionally through magneto-optical effects \cite{Dotsch2005,Catarina2020,Liu2020}, but more recently also with temporal modulations \cite{Yu2009,Lira2012,Sounas2017,Nagulu2020,Williamson2020} and in systems with a large electron mobility with current injection \cite{Duppen2016,Wenger2018,Morgado2018,Morgado2020,Dong2021,Zhao2021}. Alternatively, non-reciprocity can also be engineered using non-linear materials that are dynamically biased by the incoming wave \cite{Shadrivov2011,Fan2012,Mahmoud2015,Khanikaev2015,fernandes2018,Sounas2018,Cotrufo2021,Cotrufo2021b}. 

Systems with gain can also provide peculiar non-reciprocal responses \cite{mosfet, Buddhiraju2020}. In particular, it was recently suggested that the operation of a discrete transistor can be imitated by an hypothetical bulk nonlinear metamaterial  \cite{mosfet}. The metamaterial can be visualized as an array of discrete transistors such that the transistor response is effectively distributed over the entire space, rather than being concentrated in a lumped device as in traditional electronic systems. An hypothetical transistor-metamaterial may enable rather unique and exotic nonreciprocal and non-Hermitian optical effects (e.g., optical isolation, gainy responses, etc) \cite{mosfet}. It is thus natural to wonder if the idealized transistor-type response can be engineered with a suitable bias in a naturally available material. Remarkably, here we theoretically demonstrate that nonlinear two-dimensional systems may be the ideal platforms to implement the desired distributed transistor response. Our analysis shows that the Berry curvature dipole and unfilled electronic bands are the keys to unlock the gainy and nonreciprocal effects. Thus, the materials 
need to have a highly anisotropic electro-optical effect, typical of systems with low symmetry.

The field of nanophotonics has been ignited by the availability of novel two-dimensional (2D) materials and their unique optical properties \cite{Xia2014, Plantey2021}. Several 2D materials are promising candidates for the use in nonreciprocal devices. For instance, it has been shown that drifting electrons can break the spectral symmetry of graphene plasmons and lead to unidirectional wave propagation \cite{Duppen2016,Wenger2018,Morgado2018,Morgado2020,Dong2021,Zhao2021}. Novel 2D magnets and their heterostructures can also generate sizable magneto–optical effects \cite{Seyler2018,Henriques2020,Catarina2020} and several 2D materials can enhance nonlinear properties, including second and third order non-linearities in a wide spectral range \cite{Taghizadeh2021}. 

It is well known that some nonlinear electrical and optical effects are linked to higher order moments of the Berry curvature \cite{Sodemann2015,Xu2018,Zhang2021}. For example, the nonlinear Hall effect, which depends on the squared amplitude of the applied electric field, results from the interplay between a Berry curvature dipole (BD) and the imbalance between counterpropagating charge carriers \cite{Sodemann2015,Ma2018,Kang2019,Shi2019,He2021, NonlinearHall2021}. The circular photogalvanic effect is related to an interband Berry curvature dipole \cite{Xu2018,Kim2019}, as observed in 1Td transition metal dichalcogenides \cite{Xu2018}. 

Here we use the semiclassical Boltzmann transport theory to demonstrate that the linearized optical response of non-magnetic 2D metals is determined by the Berry curvature dipole and may mimic perfectly the linearized response of a distributed-transistor. In particular, analogous to the idealized transistor-metamaterial of Ref.  \cite{mosfet}, we find that for materials with sufficiently low-symmetry an electric static bias can lead to nonreciprocal \cite{Konig2019}  and non-Hermitian optical responses.  We apply our theoretical formalism to graphene moiré superlattices, which present  very large BD \cite{He2021,Pantaleon2021} and giant second-order nonlinear transport \cite{He2022}. We find that when the 2D-material is illuminated by a normally incident electromagnetic wave under a lateral electric bias, the ``transistor-mechanism'' can produce considerable optical gain, which depends on the light polarization. We discuss how the optical gain is influenced by the polarization of the incident wave, the Fermi energy and the number of material layers.

To begin with, we establish the relation between the linear electro-optic effect and the Berry curvature dipole in 2D metals.  Let us assume that the 2D material is biased with a static field $\mathbf{E}^0$ given by $\mathbf{E}^0=E^0_{x}\mathbf{\hat x}+E^0_{y}\mathbf{\hat y}$. For weak dynamic field variations $\mathbf{E}^\omega$ around the biasing point $\mathbf{E}(t)=\mathbf{E}^0+\mathbf{E}^{\omega}e^{-i\omega t}$, the  response is linear in both  $\mathbf{E}^\omega$ and $\mathbf{E}^0$. Semiclassically, the current density is determined by $\mathbf{J}=-e\sum_{\mathbf{k}}f_\mathbf{k}\mathbf{v}_\mathbf{k}$, where $f_\mathbf{k}$ is the electron distribution function. The band velocity  is  $\mathbf{v}_\mathbf{k}=\mathbf{v}^0_\mathbf{k}+\mathbf{v}^B_\mathbf{k}=\frac{1}{\hbar}\pd{\varepsilon_\mathbf{k}}{\mathbf{k}}+\frac{e}{\hbar}{\bf{\Omega}}_\mathbf{k}\times\mathbf{E}$, with $\mathbf{v}^B_\mathbf{k}$ being the anomalous velocity contribution determined by the Berry curvature  ${\bf{\Omega}}_\mathbf{k}$. The term  $\mathbf{v}^B_\mathbf{k}$ may be nontrivial in 2D materials with a broken inversion symmetry. To calculate the current density, we need to solve the Boltzmann transport equation (BTE) that can be written as $\pd{f_\mathbf{k}}{t}+\pd{\mathbf{k}}{t}\cdot\bm{\nabla}_\mathbf{k} f_\mathbf{k}=-\frac{\delta f_\mathbf{k}}{\tau^{sc}_\mathbf{k}}$ with $\frac{{\partial {\bf{k}}}}{{\partial t}} = \frac{{ - e}}{\hbar }{\bf{E}}$ (see the Supplementary Material [\onlinecite{SM}] for the details). The collision term is evaluated in the relaxation time approximation, where the rate of change of $f_\mathbf{k}$ is proportional to the difference $\delta f_\mathbf{k}=f_\mathbf{k}-f^0_\mathbf{k}$ between  $f_\mathbf{k}$ and the equilibrium distribution function. Here, $f^0_\mathbf{k}$ is the Fermi-Dirac distribution and $\tau^{sc}_\mathbf{k}$ is the scattering relaxation time. We want to obtain the change in $f_\mathbf{k}$ that results from the two electric fields: $\delta f_\mathbf{k}=\delta f^0_\mathbf{k}+\delta f^\omega_\mathbf{k} e^{-i\omega t}$.  The BTE is solved in two steps \cite{Cheng2014}.  Before the optical field is turned on, we consider that the DC field drives the system to a steady state.  For a DC field $\mathbf{E}^0$, the distribution function does not depend on $t$. If we assume that the change in the distribution function $\delta f^0_\mathbf{k}$ is small such that $f_\mathbf{k}\simeq f^0_\mathbf{k}+\delta f^0_\mathbf{k}$, then
$\delta f^0_\mathbf{k}\simeq \frac{e}{\hbar}\tau^{sc}_\mathbf{k}\mathbf{E}^0\cdot  \bm{\nabla}_\mathbf{k} f^0_\mathbf{k}$.  
Consider now the case where both the static and the dynamic electric fields are turned on. The response to the optical field is determined by the new steady state described by ${f_{\bf{k}}^0 + \delta f_{\bf{k}}^0}$. Thus, the BTE for the AC field is:
$-i\omega\delta f^\omega_\mathbf{k}-\frac{e}{\hbar}\mathbf{E}^\omega\cdot \bm{\nabla}_\mathbf{k} (f^0_\mathbf{k}+\delta f^0_\mathbf{k})=-\frac{\delta f^\omega_\mathbf{k}}{\tau^{sc}_\mathbf{k}}$.

The current density can be separated into AC and DC contributions $ \mathbf{J}=\mathbf{J}^0+\mathbf{J}^\omega e^{-i\omega t}$ (the second harmonic response is ignored), where the DC part $\mathbf{J}^0$ is just the linear response to the DC field. We assume that the unperturbed Hamiltonian of the 2D material is time-reversal invariant. Furthermore, only the parts of $\delta f_{\bf{k}}^0$ and $\delta f_{\bf{k}}^{\omega}$ that are odd functions of the quasi-momentum can yield a nontrivial contribution to the current. Taking this into account, it is found that the optical response is determined by the usual Drude-like linear response term (not shown) plus two contributions from the anomalous velocity, which are  linear in both static and dynamic fields. This second order contribution is known as the linear electro-optic response and can be written as $ \mathbf{J}^{\rm eo}=\mathbf{J}^{\rm eo}_{\rm H}+\mathbf{J}^{\rm eo}_{\rm NH}$ with
\begin{align}
\mathbf{J}^{\rm eo}_{\rm H}&=-\frac{e^3\tau}{\hbar^2}(\mathbf{D}\cdot\mathbf{E}^0)(\hat{\mathbf{z}}\times \mathbf{E}^\omega)=\bm{\bar{\sigma}}^{\rm eo}_{\rm H}\cdot \mathbf{E}^\omega,\label{condH}\\
\mathbf{J}^{\rm eo}_{\rm NH}&=-\frac{e^3\tau/\hbar^2}{(1-i\omega\tau)}(\hat{\mathbf{z}}\times \mathbf{E}^0)(\mathbf{D}\cdot\mathbf{E}^\omega)=\bm{\bar{\sigma}}^{\rm eo}_{\rm NH}(\omega)\cdot \mathbf{E}^\omega. \label{condNH}
\end{align}
where $\mathbf{D}$ is the Berry curvature dipole with components $D_{a}=\int\frac{d^2k}{(2\pi)^2}\Omega^z_\mathbf{k}\frac{\partial f^0_\mathbf{k}}{\partial k_a}$. In the above, $\bm{\bar{\sigma}}^{\rm eo}_{\rm H}$ and $\bm{\bar{\sigma}}^{\rm eo}_{\rm NH}(\omega)$ are linearized optical conductivities that determine the electro-optic response. Thus, $\mathbf{J}^{\rm eo}(\omega)=\bm{\bar{\sigma}}^{\rm eo}\cdot \mathbf{E}^\omega,\label{cond_eo}$ gives the piece of the dynamic current density induced by the static electric bias $\mathbf{E}^0$. In other words, the optical conductivity of the material gains an extra term $\bm{\bar{\sigma}}^{\rm eo}=\bm{\bar{\sigma}}^{\rm eo}_{\rm H}+\bm{\bar{\sigma}}^{\rm eo}_{\rm NH}$, due to the nonlinear interactions between the static and dynamical fields.

As expected, the linear electro-optic effect depends on the symmetry of the solid, and it may be  nontrivial when the inversion symmetry is broken (necessary but not sufficient condition). 
The electro-optic conductivity has two qualitatively different contributions. The first piece $\bm{\bar{\sigma}}^{\rm eo}_{\rm H}$ can be linked to an optical Hall effect \cite{Konig2019} and is associated with a gyrotropic and conservative interaction. The structure of $\bm{\bar{\sigma}}^{\rm eo}_{\rm H}$ is the same as for a lossless nonreciprocal medium with a frequency independent optical conductivity.  On the other hand, the second piece $\bm{\bar{\sigma}}^{\rm eo}_{\rm NH}$ yields a non-conservative and frequency dependent optical conductivity. Rather remarkably, the second piece can describe a process in which the optical field can extract energy from the DC field, analogous to a distributed transistor. In fact, the sign of the quadratic form $p_{\rm{dis,{\rm{EO}}}}=\frac{1}{2}{\mathop{\rm Re}\nolimits} \left\{ {{{\bf{J}}^{{\rm{eo}}}} \cdot {{\bf{E}}^{\omega ,*}}} \right\} = \frac{1}{2}{\mathop{\rm Re}\nolimits} \left\{ {{{\bf{E}}^{\omega,\ast} } \cdot \overline {\bf{\sigma }} _{{\rm{NH}}}^{{\rm{eo}}} \cdot {{\bf{E}}^{\omega}}} \right\}$ that determines the power transferred  from the optical field to the material (dissipated power) due to the linear electro-optic effect is typically unconstrained; in particular, the sign can be negative, which corresponds to optical gain. Similar to a MOSFET transistor, the optical gain arises due to the interactions between the DC current induced by the static field and the optical field. In the supplemental materials, we develop further the analogy with the transistor response.

 \begin{figure}[ht]
	\centering
	 \includegraphics[width=0.8\linewidth,clip]{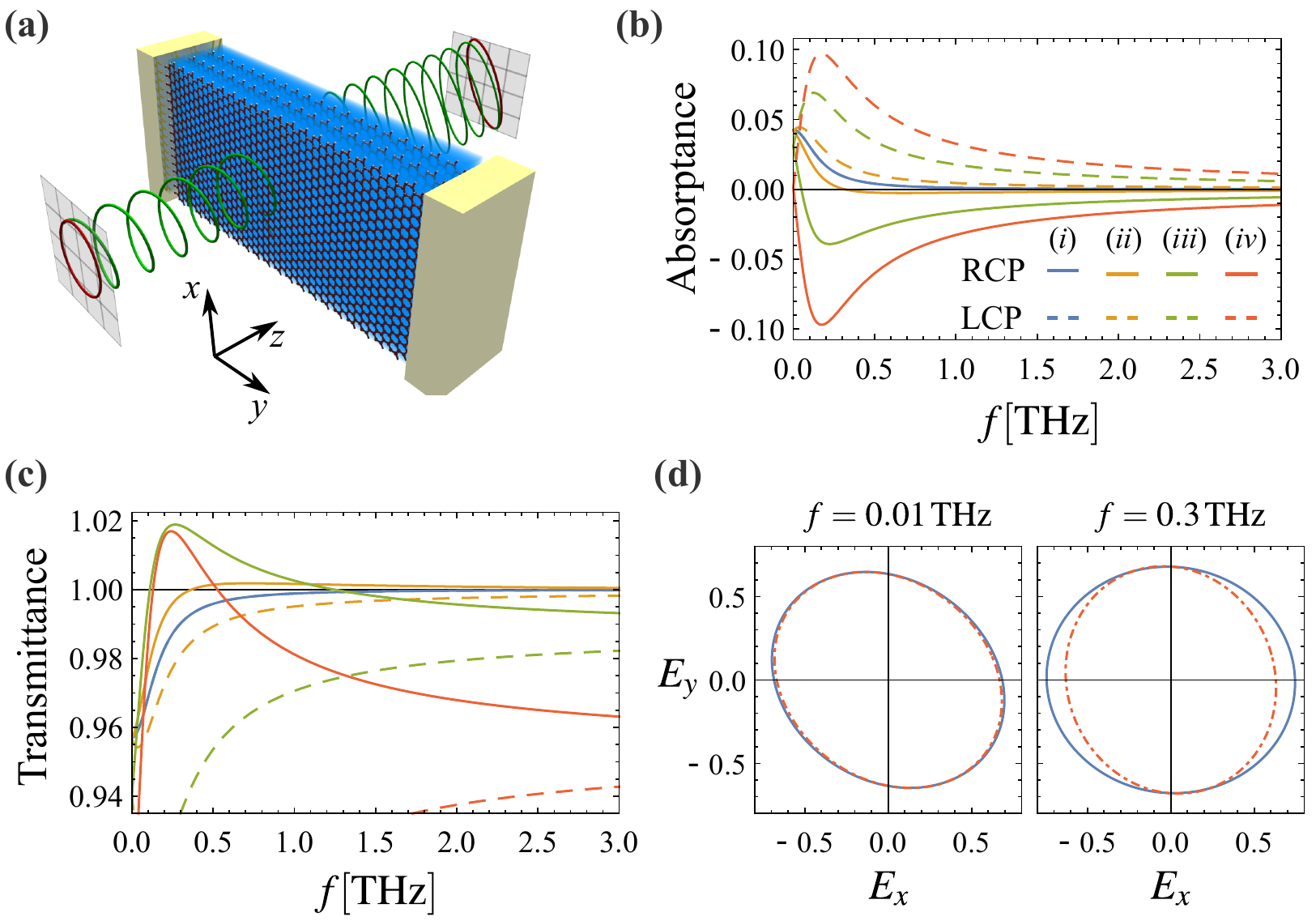}   
	\caption{(a) A 2D material described by the conductivity tensor of Eq. (\ref{optical}) is illuminated by an electromagnetic wave under normal incidence. The transmittance depends on the handedness of the incident wave. It may exceed unity for a specific handedness, while for the opposite handedness it is always less than unity. (b) Absorptance and (c) transmittance of the 2D material as a function of the frequency for an incident wave with RCP (solid) and LCP (dashed) polarization. (i) $\xi =0$; (ii) $\xi =\omega_F$; (iii) $\xi =5\omega_F$; (iv) $\xi =10\omega_F$. Note that for $\xi =0$, the absorptance (transmittance) is the same for the RCP and LCP polarizations, and thus the solid and dashed blue curves are coincident in both panels. (d) Polarization curves of the transmitted wave for an incident wave with RCP (blue solid lines) and LCP (red dashed lines) polarization for two different oscillation frequencies and  $\xi/\omega_F$=10. In the panels (b)-(d) $\omega_F/(2\pi)\simeq 0.24$ THz ($\omega_F=E_F/\hbar$ with $E_F$ =1 meV), $\gamma=1\times10^{12}$ rad/s, $\epsilon_{r1} =\epsilon_{r2}=1$.}
	\label{fig1}
\end{figure} 

Low-symmetry 2D materials are known for presenting strong optical nonlinearities \cite{Taghizadeh2021}, and thus may be the ideal platforms to observe a natural distributed-transistor response. In principle, there are several candidates for the observation of this novel effect, including, for instance, 1Td transition metal dichalcogenides such as 1Td WTe$_2$ \cite{Fei2018,Ma2018,Kang2019} or transition metal monochalcogenides \cite{BarrazaLopez2021}.  To maximize the effect, it is desirable to have a large Berry curvature dipole. The BD depends on the product of the Berry curvature and  the derivative of the distribution function. As a result, the most promising candidates for large BDs are systems with narrow gaps, which can be obtained by nanopatterning 2D materials \cite{Ho2021} or in twisted bilayers \cite{He2021,Pantaleon2021}. These systems can concentrate the band velocity and Berry curvature at the vicinity of very localized Dirac cones, increasing in this manner the BD \cite{He2021}. 

Let us first consider a generic 2D material in the metallic phase so that at low frequencies the longitudinal optical conductivity is dominated by the Drude's contribution ${{\sigma}}^{(1)}(\omega)=\sigma_D(E_F)/(\gamma-i\omega)$ with $\gamma=1/\tau$ the scattering rate.  In the case of Dirac fermions, $\sigma_D(E_F)=\sigma_0\omega_F$ where $\sigma_0=2e^2/h$ and $\omega_F=E_F/\hbar$. The Drude contribution is combined with the electro-optic conductivity $\bm{\bar{\sigma}}^{\rm eo}$. The nonconservative piece of the conductivity ($\bm{\bar{\sigma}}^{\rm eo}_{\rm NH}$) is proportional to the tensor product of the real-valued vectors $\hat{\mathbf{z}} \times \mathbf{E}^0$ and $\bf{D}$. The optical gain is maximized for high-frequencies when these two vectors are orthogonal so that the tensor $\bm{\bar{\sigma}}^{\rm eo}_{\rm NH}$ is traceless. Thus, ideally the applied static bias should be parallel to the Berry curvature dipole of the 2D material. In the following, it is assumed without loss of generality that both the Berry curvature dipole and the static bias are along the $y$-direction.
In such a case, 
the optical conductivity tensor can be written as the sum of (i) a frequency dependent part determined by both the first order optical conductivity ${{\sigma}}^{(1)}(\omega)$ and by the non-Hermitian contribution $\bm{\bar{\sigma}}^{\rm{eo}}_{\rm{NH}}(\omega)$ with (ii) a frequency independent part determined by gyrotropic optical conductivity $\bm{\bar{\sigma}}^{\rm{eo}}_{\rm{H}}$:
\begin{equation}
\bm{\bar{\sigma}}(\omega)=\frac{\sigma_0}{\gamma-i\omega}\begin{bmatrix}
\omega_F&\xi\\
0&\omega_F
\end{bmatrix}-\frac{\sigma_0}{\gamma}\begin{bmatrix}
0&-\xi\\
\xi& 0
\end{bmatrix}.
\label{optical}
\end{equation}
In the above, $\xi=\pi e D_y E_y^0/\hbar$ is a parameter with units of frequency and its sign is determined by the sign of the static bias $E_y^0$. The diagonal terms of the first piece are determined by ${{\sigma}}^{(1)}(\omega)$, whereas the anti-diagonal term of the first piece is determined by $\bm{\bar{\sigma}}^{\rm{eo}}_{\rm{NH}}(\omega)$. It should be noted that we use $\sigma^{(1)}_{xx}(\omega)=\sigma^{(1)}_{yy}(\omega)$, while the linear optical conductivity in low symmetry 2D materials is typically anisotropic. The physical mechanisms that control the optical gain depend weakly on the anisotropy, which is  ignored here for simplicity.

It can be easily checked that the total power (per unit of area) transferred from the optical field to the material is ${p_{{\rm{dis}}}} = \frac{1}{2}\frac{{{\sigma _0}}}{{{\gamma ^2} + {\omega ^2}}}\left[ {{\omega _F}\gamma \left( {{{\left| {{E_x}} \right|}^2} + {{\left| {{E_y}} \right|}^2}} \right) + \xi {\mathop{\rm Re}\nolimits} \left\{ {\left( {\gamma  + i\omega } \right)E_x^*{E_y}} \right\}} \right]$. While the first term of ${p_{{\rm{dis}}}}$ is strictly positive (corresponding to dissipation), the sign of the second term (the linear electro-optic contribution, ${p_{{\rm{dis,EO}}}}$) is unconstrained and depends mainly on the relative phase of the field components $E_x$ and $E_y$. When the second term of ${p_{{\rm{dis}}}}$ dominates, the overall material response may be ``gainy'' analogous to a conventional transistor amplifier. It should be underlined that the gainy response can be unlocked only for metallic systems, i.e. with unfilled electronic bands, so that the Berry curvature dipole and $\xi$ can be different from zero. In such a case, the static electric field induces a drift current which drives the system to a nonequilibrium steady state. In fact, without a drift current (i.e., for an insulating phase and equilibrium systems) it would be impossible to extract energy from the static voltage generator and have optical gain, in agreement with the Kleinman symmetry \cite{Kleinman1962,boyd}.

In order to study the impact of the electric bias, next we consider that the 2D material is illuminated by an electromagnetic wave that propagates along the $+z$ (normal) direction (see Figure \ref{fig1}a for a related setup with several stacked monolayers). The material is surrounded by a vacuum and is biased with a static electric field oriented along $y$. The transverse components of the reflected and transmitted waves are related to transverse components of the incident field $\mathbf{E}^{\rm{inc}}$ as  ${{\bf{E}}^{{\rm{r}}}} = {\boldsymbol{\bar \rho }} \cdot {{\bf{E}}^{{\rm{inc}}}}$ and ${{\bf{E}}^{{\rm{t}}}} = {\bf{\bar t}} \cdot {{\bf{E}}^{{\rm{inc}}}}$, where ${ \boldsymbol{\bar \rho }}$ and ${\bf{\bar t}}$ are $2 \times 2$ reflection and transmission matrices derived in the supplemental material that link the $x$ and $y$ components of the electric fields of the different waves at the interface [\onlinecite{SM}]. The transmittance $T = {\left| {{{\bf{E}}^{\rm{t}}}} \right|^2}/{\left| {{{\bf{E}}^{{\rm{inc}}}}} \right|^2}$ and the reflectance $R = {\left| {{{\bf{E}}^{\rm{r}}}} \right|^2}/{\left| {{{\bf{E}}^{{\rm{inc}}}}} \right|^2}$ can be expressed in terms of the reflection and transmission matrices as follows: $R = \left( {{{\bf{E}}^{{\rm{inc}}*}} \cdot {\bf{\bar R}} \cdot {{\bf{E}}^{\rm{inc}}}} \right)/\left( {{{\bf{E}}^{{\rm{inc}}*}} \cdot {{\bf{E}}^{{\rm{inc}}}}} \right)$ and $T = \left( {{{\bf{E}}^{{\rm{inc}}*}} \cdot {\bf{\bar T}} \cdot {{\bf{E}}^{\rm{inc}}}} \right)/\left( {{{\bf{E}}^{{\rm{inc}}*}} \cdot {{\bf{E}}^{{\rm{inc}}}}} \right)$ with ${\bf{\bar R}} = {{\boldsymbol{\bar \rho }}^\dag } \cdot {\boldsymbol{\bar \rho }}$ and ${\bf{\bar T}} = {{\bf{\bar t}}^\dag } \cdot {\bf{\bar t}}$. The $\dag$-symbol represents the operation of transpose-conjugation.

First, we consider that the incident wave is circularly polarized (CP) so that the incident electric field at $z=0^-$ is given by $\mathbf{E}^{\rm{inc}}=E_0 (\hat{\mathbf{x}}\pm i\hat{\mathbf{y}}))/\sqrt{2}$ for right (left)-handed circular polarization RCP (LCP), respectively. Figures \ref{fig1}b and \ref{fig1}c show the absorptance ($A=1-R-T$) and transmittance of the electrically-biased 2D material as a function of the frequency for different values of the ratio $\xi/\omega_F$. 
Having in mind 2D materials with minibands formed by either twisted bilayers or nano-patterning, we consider in the simulations $\omega_F/(2\pi)\sim 0.24$ THz, corresponding to Dirac fermions with $\omega_F=E_F/\hbar$, with a reduced Fermi energy $E_F$ =1 meV.  

For $\xi=0$, the absorptance for the RCP and LCP polarizations, $A_{\rm{RCP}}$ and $A_{\rm{LCP}}$ respectively, is the same because the material conductivity is isotropic. For increasing values of $\xi$, the distributed-transistor response originates optical dichroism and optical gain. The source-drain bias in 2D materials can lead to in-plane electric fields on the order of 0.2-1 V/$\mu$m \cite{Deng2020} for samples in the micrometer scale. The Berry curvature dipole $D_y$ can reach values on the order of 40 nm for graphene twisted bilayers \cite{He2021,Pantaleon2021} and 150 nm for twisted WTe$_2$ bilayers \cite{He2021};  thus, it is in principle realistic to consider $0<\xi/\omega_F<20$ for $\omega_F/(2\pi)\sim 0.24$ THz. Figure \ref{fig1}b shows that in the conservative interval $\xi/\omega_F\le 10$, there is a sizable optical gain in the terahertz range. Specifically, for RCP waves an increase of $\xi$ diminishes the material absorption and may even lead to a negative absorption (i.e., gain). Quite differently, for LCP incident waves the bias enhances the material absorption such that $1-R_{\rm{LCP}}-T_{\rm{LCP}}$ increases with $\xi$. By reversing the sign of $\xi$, which implies reversing $E^0_y$, the role of the two polarizations is interchanged (moreover, flipping the direction of arrival of the incoming wave leads to the same effect). The imbalance between the two polarizations can be attributed to both the gyrotropic and non-Hermitian parts of the electro-optic conductivity. Analogous to a lumped transistor, the optical gain is due to the energy extracted from the drifting electrons through the nonlinear response of the medium. It is important to underline that the nonreciprocal and non-Hermitian response provided by the transistor mechanism has a very different physical origin than the nonreciprocity and gain provided by current injection in graphene and related systems \cite{Duppen2016,Wenger2018,Morgado2018,Morgado2020,Dong2021,Zhao2021,Morgado2017,Morgado2021}. In fact, in the latter case the nonreciprocity and gain are rooted in the Doppler effect \cite{Morgado2018,Morgado2020} and a negative Landau damping effect \cite{Morgado2017,Morgado2021}, respectively, which require extremely large drift velocities and an ultra-high mobility. In contrast, in the transistor case it is due to the interplay of the nonlinear response and the electric bias.

The transmittance of the biased 2D material for RCP incident waves  may exceed unity ($T_{\rm{RCP}}$ > 1) over a relatively wide frequency interval (Fig. \ref{fig1}c). In contrast, the transmittance for LCP waves decreases with $\xi$ and is always less than unity ($T_{\rm{LCP}}$ < 1). The transmittance does not exhibit a monotonic behavior with $\xi$ (i.e., larger values of $\xi$ do not necessarily imply a larger transmittance, see the green and red solid curves in Fig. \ref{fig1}c). In fact, part of the energy extracted from the 2D material is coupled to the reflected wave, and thereby a negative absorptance does not imply a transmittance greater than unity \cite{SM}.

The  polarization of the incoming wave is modified by the electrically biased 2D material.  Figure \ref{fig1}d depicts the polarization curve of the transmitted waves for incident waves with RCP (blue solid lines) and LCP (red dashed lines) polarization and for two different values of the oscillation frequency. For very low frequencies, when the electro-optic response is dominated by  ${\mathop{\rm Re}\nolimits} \left\{ \bm{\bar{\sigma}}^{\rm{eo}} \right\}$, the material changes noticeably the polarization. The transmitted waves for the RCP and LCP incident waves have identical elliptical polarization with the principal axes of the ellipse parallel to $\pm\hat{\mathbf{x}}+\hat{\mathbf{y}}$. The eccentricity of the polarization curves depends on the strength of $\xi$.  For intermediate frequencies, the electro-optic conductivity is ruled by ${\mathop{\rm Im}\nolimits} \left\{ {\bm{\bar{\sigma}}^{\rm{eo}}_{\rm{NH}}} \right\}$. Even though in this case the polarization curves of the transmitted waves for incident waves with RCP and LCP polarization are different, both remain approximately circular.

By tailoring the polarization of the incident wave it is possible to further optimize the optical gain, i.e., maximize the transmittance.
The transmittance is described by the $2 \times 2$ positive definite and Hermitian matrix ${\bf{\bar T}}$ introduced previously. The transmittance is bounded by the eigenvalues ($T_{\rm{min}}$, $T_{\rm{max}}$) of ${\bf{\bar T}}$ such that: $T_{\rm{min}} \le T \le T_{\rm{max}}$. The corresponding eigenvectors $\mathbf{E}^{\rm{t}}_{\rm{eig},i}$ with $i = \left\{ {\max ,\min } \right\}$ form an orthogonal basis and determine the ``optimal polarization'' (transmittance is maximized) and the ``worst polarization'' (transmittance is minimized) of the incoming wave. It should be noted that  $\mathbf{E}^{ \rm{t}}_{\rm{eig},i}$ are not eigenvectors of $ {\bf{\bar t}}$, and thus the transmitted wave typically has a polarization that is different from that of the incident wave. The physical meaning of the eigenvectors is that they determine the polarization of the incident field that maximizes or minimizes the transmitted wave power.

\begin{figure}[htbp]
	\centering
	 \includegraphics[width=0.8\linewidth,clip]{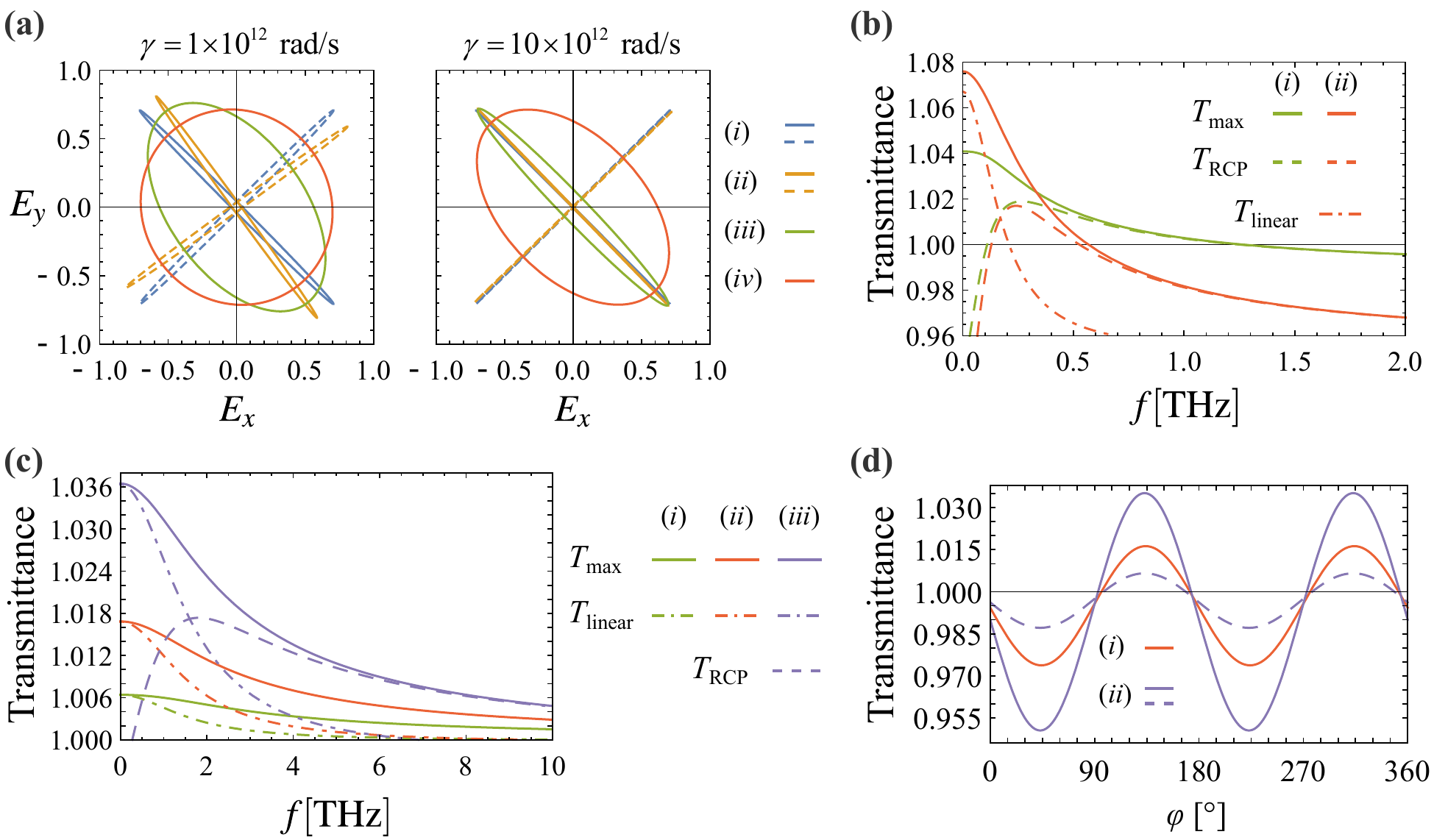}   
	\caption{ (a) Polarization curves for the incident waves that maximize (solid lines) or minimize (dashed lines) the transmittance of the 2D material for  (a) (i) $\xi =0^+$; (ii-iv) $\xi =10\omega_F$ and  (ii) $f$ =0.01THz, (iii) $f$ =0.3THz, and (iv) $f$ =3 THz; (b-c) Transmittance of the 2D material as a function of the frequency for an incident wave with optimal polarization ($T=T_{\rm{max}}$)  (solid lines), RCP polarization (dashed lines)  and linear polarization $T={T}_{\rm{linear}}$ with ${\mathbf{E}}^{\rm{inc}}=E_0 (-\hat{\mathbf{x}}+\hat{\mathbf{y}})/\sqrt{2}$ (dot-dashed lines); (b) $\gamma=1\times10^{12}$ rad/s and  (i) $\xi =5\omega_F$, (ii) $\xi =10\omega_F$. (c)  $\gamma=10\times10^{12}$ rad/s and  (i) $\xi =5\omega_F$, (ii) $\xi =10\omega_F$,  (iii) $\xi =20\omega_F$; (d) Transmittance as a function of the angle $\varphi$ for a linearly polarized incident wave with
${\mathbf{E}}^{\rm{inc}}=E_0 (\cos\varphi\hat{\mathbf{x}}+\sin\varphi\hat{\mathbf{y}})/\sqrt{2}$,  $\gamma=10\times10^{12}$ rad/s and  (i)  and for $f$ = 0.3 THz (solid lines) and $f$ = 3 THz (dashed line); (i) $\xi=10 \omega_F$. (ii) $\xi=20 \omega_F$.
The value of $\omega_F$ is the same as in Fig. \ref{fig1} for all the panels.}
	\label{fig2}
\end{figure} 

Figure \ref{fig2}a depicts the polarization curves of the incident waves that either maximize (solid lines) or minimize (dashed lines) the transmittance of the 2D material (the dashed lines are represented only for the cases $i)$ and $ii)$). In general, the  states that maximize or minimize the transmittance are elliptically polarized. The polarization states that yield $T=T_{\rm{max}}$ and $T=T_{\rm{min}}$ are rotated by $90^{\circ}$. For low frequencies, $T=T_{\rm{max}}$ corresponds to an incident wave that is almost linearly polarized with an orientation of $135^{\circ}$  with respect to the +$x$-axis. If the scattering rate is higher, the peak of the real part of $\bm{\bar{\sigma}}^{\rm{eo}}_{\rm{NH}}$ broadens. This increases the eccentricity of the polarization curve that yields $T=T_{\rm{max}}$. For larger frequencies, the eccentricity decreases and the optimal polarization tends to the RCP polarization. 

This behavior can be understood by analyzing the component of the gain ($-p_{\rm{dis,EO}}$) due to the electro-optic response: $\frac{1}{2}\frac{{{-\sigma _0 \xi}}}{{{\gamma ^2} + {\omega ^2}}} {  {\mathop{\rm Re}\nolimits} \left\{ {\left( {\gamma  + i\omega } \right) E_x^*{E_y}} \right\}}$. For low frequencies, the gain per unit of area is roughly $ \frac{1}{2}\frac{{{\sigma _0 \xi \gamma}}}{{{\gamma ^2} + {\omega ^2}}}   {\mathop{\rm Re}\nolimits} \left\{ {-E_x^*{E_y}} \right\}$, which for $\xi>0$ and a fixed incident power flux is maximized (negative dissipation) when $E_x$ and $E_y$ have the same amplitude and are in opposition of phase (linear polarization). On the other hand, for large frequencies the gain is approximately $ \frac{1}{2}\frac{{{\sigma _0 \xi \omega}}}{{{\gamma ^2} + {\omega ^2}}} {  {\mathop{\rm Re}\nolimits} \left\{ {-i E_x^*{E_y}} \right\}} $, which is maximized when $E_x$ and $E_y$ have the same amplitude and are in quadrature (circular polarization).

Figure \ref{fig2}b compares the transmittance of the state that yields $T=T_{\rm{max}}$ with the transmittance for a RCP wave and for a linearly polarized wave with  ${\mathbf{E}}^{\rm{inc}} = E_0 (-\hat{\mathbf{x}}+\hat{\mathbf{y}})/\sqrt{2}$. The state with optimal polarization provides considerably more optical gain than the RCP state in the low frequency regime, especially for a large $\xi/\omega_F$. On the other hand, the transmittance of the linearly polarized wave approaches $T_{\rm{max}}$ in the limit $\omega\rightarrow 0$, which is consistent with the analysis of Figure \ref{fig2}a. However, for intermediate frequencies, $T_{\rm{linear}}$ is strongly reduced. Figure \ref{fig2}d shows the transmittance for an incoming wave with linear polarization as a function of the orientation of the electric field vector. Consistent with the previous discussion, the transmittance is maximized for $\varphi = 135^{\circ}$. 

In general, a lower scattering rate leads to a stronger gain due to the larger mobility of electrons (compare Fig. \ref{fig2}b with Fig. \ref{fig2}c, which was calculated with a scattering rate 10 times larger). However, for sufficiently small $\gamma$ a non-monotonic behavior is observed due to an impedance mismatch effect that repels the electric field lines away from the surface of the 2D material and suppresses the gain (not shown). The impedance mismatch is mostly due to the gyrotropic part of the material response. In principle it is possible to counteract the undesired effects of a large scattering rate with some optimization of the Berry curvature dipole (i.e., using a larger $\xi/\omega_F$, see Fig. \ref{fig2}c). This is made possible by the broad peak in
${\mathop{\rm Im}\nolimits} \left\{ {\bm{\bar{\sigma}}^{\rm{eo}}_{\rm{NH}}} \right\}$.

\begin{figure}[htbp]
	\centering
	 \includegraphics[width=0.99\linewidth,clip]{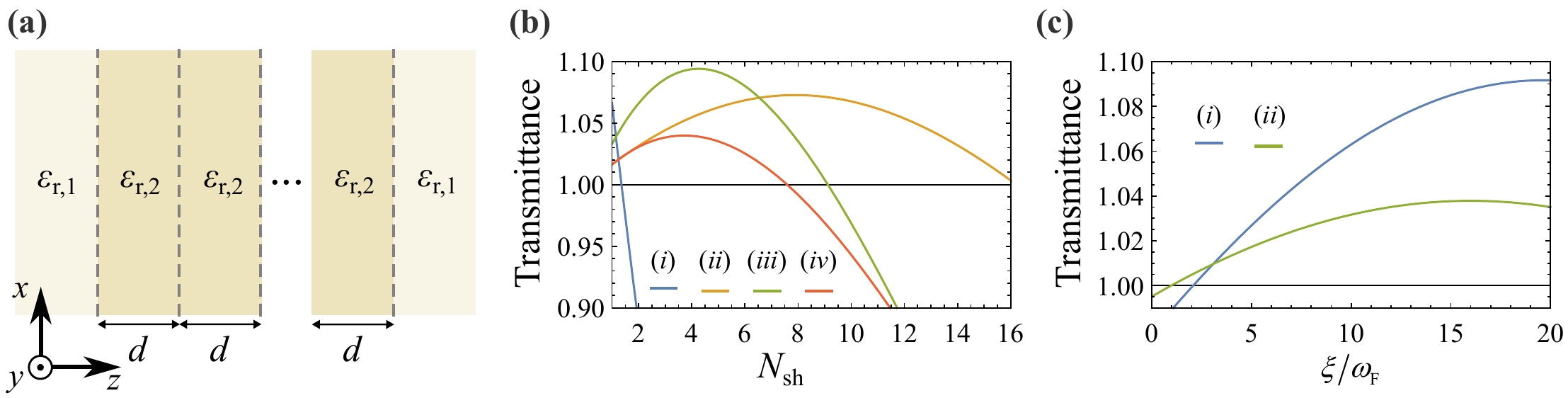}   
	\caption{(a) Geometry of a heterostructure formed by a stack of 2D materials separated
by thin dielectrics. (b-c) Transmittance of the multilayer system for an incident wave with the optimal polarization ($T=T_{\rm{max}}$), $E_F$ = 1 meV, $\epsilon_{r,1}=1$,  $\epsilon_{r,2}=3.5$ (hBN), and $d$ = 10 nm. (b) Transmittance as a function of the number of sheets $N_{\rm{sh}}$ for (i) $\gamma=1\times10^{12}$ rad/s, $\xi =10\omega_F$ and $f$ = 0.3 THz, (ii-iv) $\gamma=10\times10^{12}$ rad/s, (ii) $\xi =10\omega_F$ and $f$ = 0.3 THz, (iii) $\xi =20\omega_F$ and $f$ = 0.3 THz, and (iv) $\xi =20\omega_F$ and $f$ = 3 THz. (c) Transmittance of the multilayer system as a function of $\xi/\omega_F$ for $N_{\rm{sh}}$=5, $\gamma=10\times10^{12}$ rad/s, and (i) $f$ = 0.3 THz, (ii) $f$ = 3 THz.}
 \label{fig3}
\end{figure}

The optical gain can be boosted by stacking several layers of the 2D material. To illustrate this, next we consider a heterostructure formed by $N_{\rm{sh}}$ identical 2D material sheets separated by dielectric spacers of thickness $d$ and dielectric constant $\epsilon_{r2}$ surrounded by a vacuum (see Figure \ref{fig3}a).  Using a transfer matrix formalism, the transverse electromagnetic fields at the input interface ($z=0^-$) can be linked to the fields at the  output interface  ($z=(N-1)d^+$). In this manner, one can find the transmittance of the heterostructure as a function of the polarization state of the incoming wave \cite{SM}.
Figure \ref{fig3}b shows  $T_{\rm{max}}$ as a function of the number of layers for a heterostructure with a spacer with $d=$10 nm and $\epsilon_{r2}=3.5$, which is the dielectric constant of hBN \cite{Dean2010}. The transmittance for the optimal polarization can reach 110$\%$. As seen, the transmittance reaches a maximum for some optimum $N_{\rm{opt}}$ number of layers that depends on the dielectric spacers, on the static field bias and on the scattering rate.  

 The origin of the optimum $N_{\rm{opt}}$ is the change of the wave polarization caused by the transmission through each layer. In fact, the wave that illuminates a certain layer has typically a sub-optimal polarization. This problem may be fixed with polarization transformers that restore the optimal polarization after the transmission through each layer.
 
 A larger scattering rate and smaller $\xi$ reduce the polarization rotation suffered by the transmitted waves, increasing the value of $N_{\rm{opt}}$. Figure \ref{fig3}c presents the transmittance of a heterostructure with $N_{\rm{sh}}=5$ layers as a function of $\xi/\omega_F$ (similar results are obtained for other values of $N_{\rm{sh}}$). As seen, the transmittance saturates for some value of $\xi/\omega_F$ and then drops down. This property is due to the impedance mismatch caused by the gyrotropic part of the electro-optic conductivity  $\bm{\bar{\sigma}}^{\rm{eo}}_{\rm{H}}$, which is also proportional to $\xi$. In fact, without this component the transmittance would increase monotonically with $\xi/\omega_F$.
 
Next, we analyze the possibility of observing the  distributed-transistor response in twisted bilayer graphene (TBG). It is well known that ``strain'' can be used to tune the Berry curvature dipole in 2D materials \cite{You2018,Batillomo2019,Son2019,Ho2021,Arora2021}. Deformations of  0.1 - 0.5$\%$ have been measured with STM in TBG \cite{Kerelsky2019,Xie2019,Kazmierczak2021}. Consequently, TBG encapsulated in hBN naturally presents a broken inversion symmetry \cite{Hunt2013,Finney2019} and heterostrain \cite{Kerelsky2019,Xie2019,Kazmierczak2021}, which also breaks the ${\cal C}_3$ symmetry. Both effects reduce the symmetry of the TBG and allow for the presence of a Berry curvature dipole. In fact, the combination of uniaxial strain and mini-gaps in TBG leads to extremely large BDs that can reach values on the order of 10$^2$ nm \cite{He2021,Pantaleon2021,Arora2021}. Thus, TBG is a promising candidate for observing the current-driven optical gain. 

 We follow a standard theoretical approach to obtain the BD and the optical conductivity tensor of strained TBG for strain in the range of $0.1\%<\epsilon<0.3\%$ \cite{He2021,Pantaleon2021}. We model the heterostructure with a low-energy continuum Hamiltonian consisting of two Dirac terms for the graphene layers and a tunneling term for the hopping between layers \cite{Bistritzer2011}. The key ingredients to observe optical gain are a large BD and small longitudinal conductivity, such that the off-diagonal terms of the conductivity tensor dominate the optical response. This can be achieved with twist angles on the order of $1.25^\circ$. For these angles,  the maximum value of the BD occurs near the bottom of the first conduction moir\'e band, where the longitudinal conductivity is small \cite{SM}.  

\begin{figure}[htbp]
	\centering
	 \includegraphics[width=0.99\linewidth,clip]{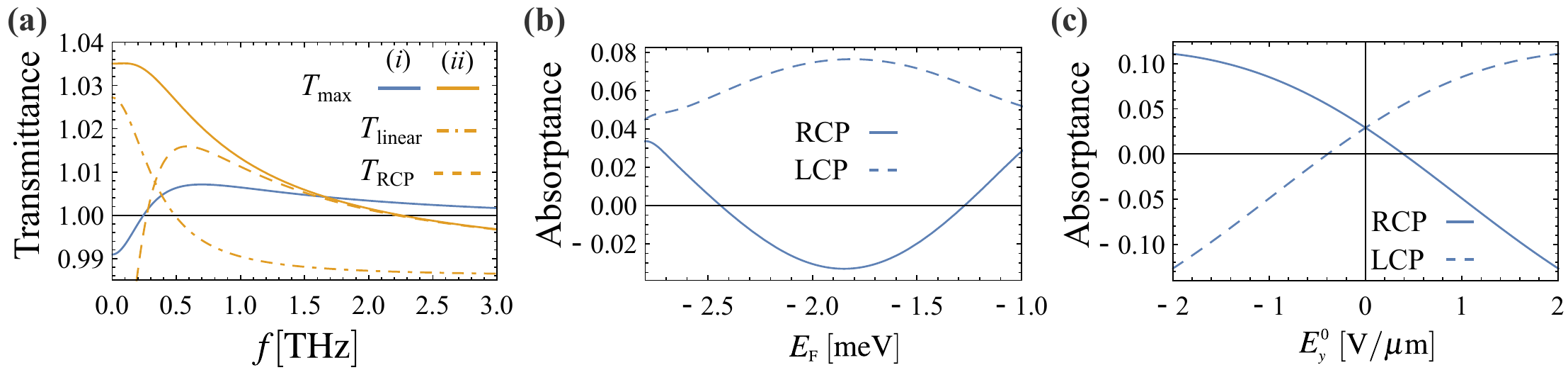}   
	\caption{(a) Transmittance and (b-c) absorptance of the strained twisted bilayer graphene (surrounded by a vacuum) for an incident wave with optimal polarization ($T=T_{\rm{max}}$) (solid lines),  RCP polarization ($T=T_{\rm{RCP}}$) (dashed lines), and linear polarization ($T=T_{\rm{linear}}$) with ${\mathbf{E}}^{\rm{inc}}=E_0 (-\hat{\mathbf{x}}+\hat{\mathbf{y}})/\sqrt{2}$ (dot-dashed lines), for $\gamma=2\times 10^{12}$ rad/s. (a) Transmittance as a function of the frequency for $E_y^0 $ = 0.8  V/$\mu$m; (i) $E_F$ = -2.5 meV, (ii) $E_F$ = -1.8 meV. (b) Absorptance
as a function of  $E_F$ for $E_y^0 $ = 0.8  V/$\mu$m and $f$ = 0.3 THz. (c) Absorptance as a function of amplitude of the static bias field directed along the $y$ direction for $E_F$ = -1.8 meV and $f$ = 0.3 THz.}
	\label{fig4}
\end{figure} 

Figure \ref{fig4}a depicts $T_{\rm{max}}$, $T_{\rm{RCP}}$, and $T_{\rm{linear}}$ for electrically biased TBG as a function of the frequency for $E_y^0 =0.8$ V/$\mu$m, $\gamma=2\times 10^{12} \rm{rad/s}$ (equivalent to a broadening of 1.6 meV) and two different values of $E_F$, specifically (i) $E_F=-2.5$ meV and (ii) $E_F=-1.8$ meV. The 2D material can provide considerable gain over a wide range of frequencies. The transmittance exhibits a behavior analogous to the previous figures. For low frequencies, the optimal polarization is linear, whereas for high frequencies the optimal polarization becomes circular. The transmittance depends significantly on $E_F$, due to the reasons discussed next.

The value of $E_F$ affects the optical gain through changes in $D_y$ and in the longitudinal conductivity. In fact, it is well known that the Berry curvature dipole is strongly dependent on $E_F$ in TBG \cite{Batillomo2019,Son2019,Ho2021,Arora2021}. Consistent with previous works, we show in the supplementary materials that $D_y$ is a non-monotonical function of the Fermi energy [\onlinecite{SM}]. On the other hand, the Fermi energy also controls the longitudinal conductivity which should be kept as small as possible to maximize $\xi/\omega_F$. 
As illustrated in Fig. \ref{fig4}b, values of $E_F$ near the bottom of the conduction band ($E_F\simeq -2.5$ meV) provide a good compromise between a large dipole $D_y$ and a small longitudinal conductivity. The negative absorptance is maximized for $E_F=-1.8$  meV where the BD has contributions from both electron and hole pockets [\onlinecite{SM}]. It should be noted that in an experiment it may be easier to detect the active response in terms of positive and negative variations of the absorption.

Figure \ref{fig4}c shows how the absorptance varies with the static bias. The material exhibits a ``gainy'' response for a wide range of values of the bias field. The change in the sign of the absorptance with the sign of the bias and its sensitivity to the handedness of the wave polarization are the ``smoking gun'' signatures of the transistor-like optical response. 

In summary, our theoretical analysis reveals that electrically biased two-dimensional metallic systems with a large Berry curvature dipole may effectively behave as ``distributed transistors'' with a strongly nonreciprocal and non-Hermitian distributed optical response. The described effect is rooted in a novel electro-optic effect that arises due to the interplay of the drift current induced by the static bias and material nonlinearities. 
In order to illustrate the rather unique properties of the material, we characterized the absorptance and transmittance of the electrically biased 2D material under plane wave illumination. Our analysis reveals that the material response can be either gainy or lossy depending on the handedness of the polarization state of the incident wave. The optical gain/loss also depends on the direction of the wave propagation and on the orientation of the static electric field, giving the flexibility to design of a panoply of new tunable and active devices (e.g., amplifiers, oscillators, etc) with the gain controlled by the polarization of the wave. Furthermore, it was shown that an heterostructure formed by multiple material layers can boost the optical gain. Finally, we identified graphene moiré supperlattices as very promising platforms to observe the transistor-like electro-optic effect.  
\begin{acknowledgements}
This work was partially funded by  the Institution of Engineering and Technology (IET) under the A F Harvey Research Prize 2018, by the Simons Foundation under the award 733700 (Simons Collaboration in Mathematics and Physics, ”Harnessing Universal Symmetry Concepts for Extreme Wave Phenomena”) and by Instituto de Telecomunicações under Project Number UID/EEA/50008/2020.
T. A. M. acknowledges FCT for research financial support with reference CEECIND/04530/2017 under the CEEC Individual 2017, and IT-Coimbra for the contract as an assistant researcher with reference CT/N◦. 004/2019- F00069.
S.L. acknowledges FCT and IT-Coimbra for the research financial support with reference DL 57/2016/CP1353/CT000.  
\end{acknowledgements}

\bibliographystyle{apsrev}

%
\onecolumngrid
\newpage
\appendix
\renewcommand{\theequation}{S\arabic{equation}}
\renewcommand{\thefigure}{S\arabic{figure}}
\setcounter{page}{1}
\setcounter{equation}{0}
\setcounter{figure}{0}

\begin{center}
{\bf Supplementary material for ``Engineering transistor-like optical gain in two-dimensional materials with Berry curvature dipoles''}	
\end{center}

\section{A. Boltzmann Transport Equation (BTE)}

 Let us assume that the 2D material is biased with a static field $\mathbf{E}^0$ given by $$\mathbf{E}^0=E^0_{x}\mathbf{\hat x}+E^0_{y}\mathbf{\hat y}.$$ 
 For weak dynamic field variations $\mathbf{E}^\omega$ around the biasing point $\mathbf{E}(t)=\mathbf{E}^0+\mathbf{E}^{\omega}e^{-i\omega t}$, the  response is linear in both  $\mathbf{E}^\omega$ and $\mathbf{E}^0$. Semiclassically, the current density is determined by $$\mathbf{J}=-e\sum_{\mathbf{k}}f_\mathbf{k}\mathbf{v}_\mathbf{k},$$
where $f_\mathbf{k}$ is the electron distribution function. The band velocity  is  $$\mathbf{v}_\mathbf{k}=\mathbf{v}^0_\mathbf{k}+\mathbf{v}^B_\mathbf{k}=\frac{1}{\hbar}\pd{\varepsilon_\mathbf{k}}{\mathbf{k}}+\frac{e}{\hbar}{\bf{\Omega}}_\mathbf{k}\times\mathbf{E},$$ with $\mathbf{v}^B_\mathbf{k}$ being the anomalous velocity contribution determined by the Berry curvature  ${\bf{\Omega}}_\mathbf{k}$. 

To calculate the current density, we need to solve the Boltzmann transport equation (BTE)
\begin{equation}
\pd{f_\mathbf{k}}{t}+\pd{\mathbf{k}}{t}\cdot\bm{\nabla}_\mathbf{k} f_\mathbf{k}=-\frac{\delta f_\mathbf{k}}{\tau^{sc}_\mathbf{k}}, ~~ \mbox{   with  } ~~\frac{{\partial {\bf{k}}}}{{\partial t}} = \frac{{ - e}}{\hbar }{\bf{E}}. 
\end{equation}
The collision term is evaluated in the relaxation time approximation, where the rate of change of $f_\mathbf{k}$ is proportional to the difference $\delta f_\mathbf{k}=f_\mathbf{k}-f^0_\mathbf{k}$ between  $f_\mathbf{k}$ and the equilibrium distribution function. Here, $f^0_\mathbf{k}$ is the Fermi-Dirac distribution and $\tau^{sc}_\mathbf{k}$ is the scattering relaxation time. We want to obtain the change in $f_\mathbf{k}$ that results from the two electric fields: $\delta f_\mathbf{k}=\delta f^0_\mathbf{k}+\delta f^\omega_\mathbf{k} e^{-i\omega t}$.  The BTE is solved in two steps \cite{Cheng2014}.  Before the optical field is turned on, we consider that the DC field drives the system to a steady state.  For a DC field $\mathbf{E}^0$, the distribution function does not depend on $t$. If we assume that the change in the distribution function $\delta f^0_\mathbf{k}$ is small such that $f_\mathbf{k}\simeq f^0_\mathbf{k}+\delta f^0_\mathbf{k}$, then
$$\delta f^0_\mathbf{k}\simeq \frac{e}{\hbar}\tau^{sc}_\mathbf{k}\mathbf{E}^0\cdot  \bm{\nabla}_\mathbf{k} f^0_\mathbf{k}.$$  
Consider now the case where both the static and the dynamic electric fields are turned on. The response to the optical field is determined by the new steady state described by ${f_{\bf{k}}^0 + \delta f_{\bf{k}}^0}$. Thus, the BTE for the AC field is:
$$-i\omega\delta f^\omega_\mathbf{k}-\frac{e}{\hbar}\mathbf{E}^\omega\cdot \bm{\nabla}_\mathbf{k} (f^0_\mathbf{k}+\delta f^0_\mathbf{k})=-\frac{\delta f^\omega_\mathbf{k}}{\tau^{sc}_\mathbf{k}}.$$
The solution can be written explicitly as:
 \begin{equation}
 \delta f^\omega_\mathbf{k}\simeq \frac{e\tau/\hbar}{(1-i\omega\tau)}\mathbf{E}^\omega\cdot \bm{\nabla}_\mathbf{k} f^0_\mathbf{k} +\frac{(e\tau/\hbar)^2}{(1-i\omega\tau)}\mathbf{E}^\omega\cdot \bm{\nabla}_\mathbf{k}(\mathbf{E}^0\cdot  \bm{\nabla}_\mathbf{k} f^0_\mathbf{k}),
 \label{ACd}
 \end{equation}
where $\tau^{sc}_\mathbf{k}\sim \tau$ is assumed a constant. The first term is linear in $\mathbf{E}^\omega$ while the second is quadratic in the electric field as it depends on both $\mathbf{E}^0$ and  $\mathbf{E}^\omega$. The induced current density can be written in terms of the total distribution function as follows:
\begin{equation}\mathbf{J}=-e\sum_\mathbf{k}(f^0_\mathbf{k}+\delta f^0_\mathbf{k}+\delta f^\omega_\mathbf{k} e^{-i\omega t} )(\mathbf{v}^0_\mathbf{k}+\frac{e}{\hbar}{\bf{\Omega}}_\mathbf{k} \times (\mathbf{E}^0+\mathbf{E}^\omega e^{-i\omega t})).
\end{equation}
This expression can be separated into AC and DC contributions $ \mathbf{J}=\mathbf{J}^0+\mathbf{J}^\omega e^{-i\omega t}$ (the second harmonic response is ignored), where the DC part $\mathbf{J}^0$ is just the linear response to the DC field. We assume that the unperturbed Hamiltonian of the 2D material is time-reversal invariant, and thereby the term $f^0_\mathbf{k}{\bf{\Omega}}_\mathbf{k}\times \mathbf{E}^\omega$ does not contribute after integration over the momentum space. Furthermore, only the parts of $\delta f_{\bf{k}}^0$ and $\delta f_{\bf{k}}^{\omega}$ that are odd functions of the quasi-momentum can yield a nontrivial contribution to the current. Taking this into account, it is found that the optical response is determined by the usual Drude-like linear response term (not shown) plus two contributions from the anomalous velocity:
\begin{equation}
\mathbf{J}^{\rm eo}=-e\sum_\mathbf{k}\delta f^0_\mathbf{k}\frac{e}{\hbar}({\bf{\Omega}}_\mathbf{k}\times \mathbf{E}^\omega)-e\sum_\mathbf{k}\delta f^{\omega0}_\mathbf{k}\frac{e}{\hbar}({\bf{\Omega}}_\mathbf{k}\times \mathbf{E}^0).
\end{equation}
with $\delta f_{\bf{k}}^{\omega0} = \frac{1}{{1 - i\omega \tau }}\frac{{\tau e}}{\hbar }{\nabla _{\bf{k}}}f_{\bf{k}}^0 \cdot {{\bf{E}}^\omega }$.

 In 2D materials the Berry curvature is of the form $\boldsymbol{\Omega}_\mathbf{k} =\Omega_\mathbf{k}^z \hat{\mathbf{z}}$. Hence, the second order contribution can be written as $ \mathbf{J}^{\rm eo}=\mathbf{J}^{\rm eo}_{\rm H}+\mathbf{J}^{\rm eo}_{\rm NH}$, as shown in the main text.


\section{B. Reflection and transmission matrices for a 2-dimensional material}
\begin{figure}[htbp]
	\centering
	 \includegraphics[width=0.55\linewidth,clip]{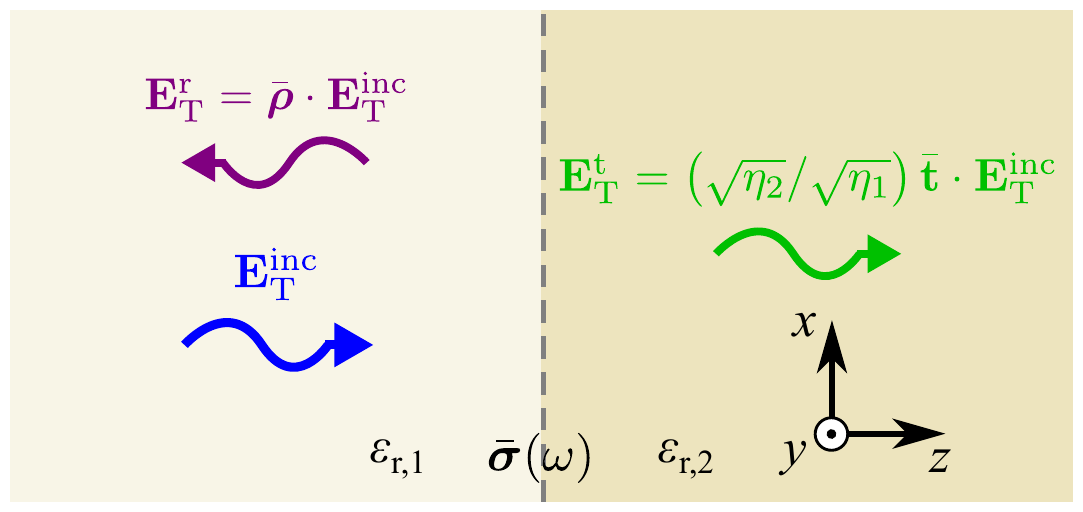}   
	\caption{Sketch of a 2D material surrounded by two dielectrics with relative permittivities $\epsilon_{r,1}$ and $\epsilon_{r,1}$ and wave impedances $\eta_1 =\eta_0/\sqrt{\epsilon_{r,1}}$ and $\eta_2 =\eta_0/\sqrt{\epsilon_{r,2}}$.} 
	\label{figs1}
\end{figure}

Here we derive the reflection and transmission matrices (${\boldsymbol{\bar \rho }}$ and  ${\bf{\bar t }}$) for a 2-dimensional (2D) material whose electromagnetic response is determined by the surface conductivity $\bm{\bar{\sigma}}(\omega)$ (Fig. \ref{figs1}).
Following the approach of Refs. \cite{Morgado2016,Latioui2019}, we define the transverse fields as:
\begin{equation}
\mathbf{E}_{\rm T} =\begin{bmatrix}
           E_x \\
           E_y
         \end{bmatrix}, \quad   \mathbf{\bar{J}}\cdot\mathbf{H}_{\rm T}  =\begin{bmatrix}
           0 & 1\\
           -1 &0
         \end{bmatrix}
         \begin{bmatrix}
           H_x \\
           H_y
 \end{bmatrix}
 =
 \begin{bmatrix}
           H_y \\
           -H_x
 \end{bmatrix}
\end{equation}

The transverse components of the electric $\mathbf{E} $ and magnetic $\mathbf{H} $ fields in the dielectrics $i=1,2$ are linked by
\begin{equation}
\mathbf{\bar{J}}\cdot\mathbf{H} ^\pm_{\rm T}=\pm\mathbf{\bar{Y}}_i\cdot\mathbf{E} ^\pm_{\rm T}
\end{equation}
where the sign $\pm$ determines if the wave propagates towards the $+z$ or the $-z$ direction and
\begin{equation}
\mathbf{\bar{Y}}_i= \frac{1}{\eta_0k_0k_{z,i}}\begin{bmatrix}
           k_0^2\epsilon_{r,i}-k_y^2 & k_xk_y\\
           k_xk_y &k_0^2\epsilon_{r,i}-k_x^2
         \end{bmatrix}        
\end{equation}
where $k_{z,i}=\sqrt{k_0^2\epsilon_{r,i}-k_x^2 -k_y^2}$, $k_0=\omega/c$ is the free-space wave number, and $\eta_0$ is the vacuum impedance.
By matching the tangential component
of the electric field ($\mathbf{E}_{\rm T} |_{z=0^+}-\mathbf{E}_{\rm T} |_{z=0^-}=0$) and by imposing the surface impedance boundary condition  ($-\mathbf{\bar{J}}\cdot(\mathbf{H}_{\rm T} |_{z=0^+}-\mathbf{H}_{\rm T} |_{z=0^-})=\bm{\bar{\sigma}}\cdot\mathbf{E}_{\rm T}  $) at the interface, it is found that:
\begin{eqnarray}
&(\mathbf{1}_{2\times2}+{\boldsymbol{\bar \rho }})\cdot\mathbf{E} ^{\rm{inc}}_{\rm T}=\mathbf{E} ^{\rm{t}}_{\rm T}&\label{S4}\\
&\mathbf{\bar{Y}}_1\cdot \mathbf{E} ^{\rm{inc}}_{\rm T}-\mathbf{\bar{Y}}_1\cdot\mathbf{E} ^{\rm{r}}_{\rm T}-\mathbf{\bar{Y}}_2\cdot\mathbf{E} ^{\rm{t}}_{\rm T}=\bm{\bar{\sigma}}\cdot\mathbf{E} ^{\rm{t}}_{\rm T} \label{S5}&
\end{eqnarray}
The reflected and transmitted fields are related to the incident electric field as shown in Fig. \ref{figs1}.
From Eqs. \eqref{S4} and \eqref{S5} we obtain the following reflection and transmission matrices:
\begin{equation}
{\boldsymbol{\bar \rho }}=\begin{bmatrix}
           \rho_{xx} & \rho_{xy}\\
           \rho_{yx} & \rho_{yy}
         \end{bmatrix}=(\mathbf{\bar{Y}}_1+\mathbf{\bar{Y}}_2+\bm{\bar{\sigma}})^{-1}\cdot(\mathbf{\bar{Y}}_1-\mathbf{\bar{Y}}_2-\bm{\bar{\sigma}})\
         \label{S6}
\end{equation}
\begin{equation}
{\bf{\bar t }}=\begin{bmatrix}
           t_{xx} & t_{xy}\\
           t_{yx} & t_{yy}
         \end{bmatrix}=2\sqrt{\frac{\eta_1}{\eta_2}}(\mathbf{\bar{Y}}_1+\mathbf{\bar{Y}}_2+\bm{\bar{\sigma}})^{-1}\cdot\mathbf{\bar{Y}}_1\
         \label{S7}
\end{equation}
where the generic $lm$ element represents the complex amplitude of the reflected ($\rho$) or transmitted ($t$) wave, linearly polarized in the $l$ direction for the excitation in the $m$ direction.

 \section{C. Study of the reflectance} 
 
\begin{figure}[h]
	\centering
	 \includegraphics[width=0.45\linewidth,clip]{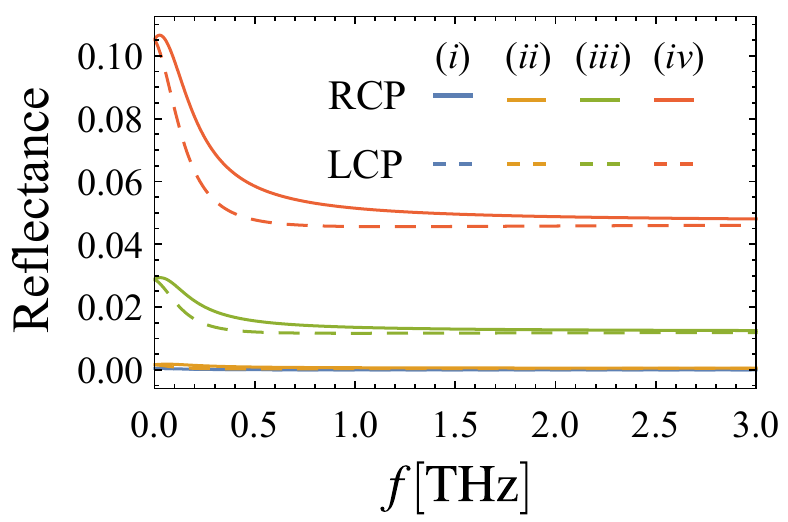}   
	\caption{Reflectance of the 2D material as a function of the frequency for an incident wave with RCP (solid) and LCP (dashed) polarization, for $\omega_F/(2\pi)\simeq 0.24$ THz ($\omega_F=E_F/\hbar$ with $E_F$ =1 meV), $\gamma=1\times10^{12}$ rad/s, $\epsilon_{r1} =\epsilon_{r2}=1$. (i) $\xi =0$; (ii) $\xi =\omega_F$; (iii) $\xi =5\omega_F$; (iv) $\xi =10\omega_F$. Note that for $\xi =0$, the reflectance  is the same for the RCP and LCP polarizations, and thus the solid and dashed blue curves are coincident.} 
	\label{figs3}
\end{figure}

 Figure \ref{figs3} depicts the reflectance of the biased 2D material for the same parameters as in Fig. 1 of the main text as a function of the frequency for RCP (solid lines) and LCP (dashed lines) incident waves. The reflectance is weakly dependent on the incident wave polarization and typically increases with $\xi$, which explains why a transmittance less than unity can be compatible with a negative absorption for large values of $\xi$.
 
\newpage

\section{D. Reflection and transmission matrices for a multilayer system}
\begin{figure}[htbp]
	\centering
	 \includegraphics[width=0.65\linewidth,clip]{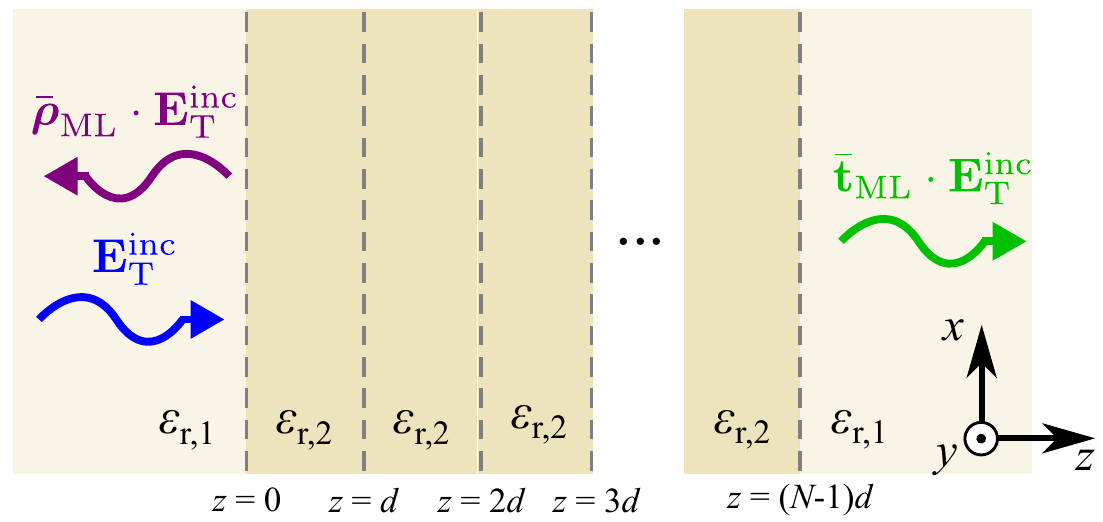}   
	\caption{Sketch of a multilayer system formed by $N$ 2D material sheets.} 
	\label{figs2}
\end{figure}

Here, we derive the transmission matrix for a multilayer structure formed by several 2D material sheets and dielectric layers (Fig. \ref{figs2}). 

The boundary conditions that link the electromagnetic fields across the 2D material (see section A) interfaces can be written in a compact matrix form as:
\begin{equation}
\begin{bmatrix}
           E_x \\
           E_y \\
           H_x \\
           H_y 
  \end{bmatrix}_{z=z_0^+} = \mathbf{\bar{U}}_B\cdot \begin{bmatrix}
           E_x \\
           E_y \\
           H_x \\
           H_y 
  \end{bmatrix}_{z=z_0^-}
     \mbox{with \,\,}   \mathbf{\bar{U}}_B=
  \begin{bmatrix}
           1 & 0 & 0 & 0\\
           0 & 1 & 0 & 0 \\
           \sigma_{yx} & \sigma_{yy} &1&0 \\
            -\sigma_{xx} & -\sigma_{xy} &0&1 \\
         \end{bmatrix}.\label{S13}
\end{equation}
$\mathbf{\bar{U}}_B$ is a transmission matrix that relates the transverse fields at $z_0^+$ with the fields calculated at  $z_0^-$.

To characterize the propagation in the dielectric slabs, we rely on the Maxwell's equations:
\begin{equation}
\vec{\nabla}\times\mathbf{E}=i\omega\mu_0\mathbf{H},\quad   \vec{\nabla}\times\mathbf{H}=-i\omega\epsilon_0\epsilon_{r,i}\mathbf{E}.
\end{equation}
The material is assumed infinite along the $x$ and $y$ directions. In this case the fields can be decomposed into plane waves with spatial variation $e^{i(k_xx+k_yy)}$ and the nabla operator reduces to $\vec{\nabla}= ik_x\hat{\mathbf{x}}+ik_y\hat{\mathbf{y}}+\partial_z \hat{\mathbf{z}}$. Thus, we can write:
\begin{equation}
-i\begin{bmatrix}
           ik_yE_z-\partial_zE_y \\
           \partial_zE_x-ik_xE_z \\
           ik_xE_y-ik_yE_x
          \end{bmatrix}=\omega\mu_0\mathbf{H}
    \mbox{\,\,and\quad}   
 i\begin{bmatrix}
           ik_yH_z-\partial_zH_y \\
           \partial_zH_x-ik_xH_z \\
           ik_xH_y-ik_yH_x
          \end{bmatrix}=\omega\epsilon_0\epsilon_{r,i}\mathbf{E}.
 \end{equation}
The $z$-component of the fields does not depend on $\partial_z$  and can be expressed as a function of the other components
\begin{equation}
E_z=\frac{-k_xH_y+k_yH_x}{\omega\epsilon_0\epsilon_{r,i}}
    \mbox{\,\,and\quad}   
H_z=\frac{k_x E_y-k_y E_x}{\omega\mu_0.}
\end{equation}
From here, it follows that the remaining set of equations can be written in a matrix form as: 
\begin{equation}
i\partial_z\begin{bmatrix}
           E_x \\
           E_y \\
           H_x \\
           H_y 
  \end{bmatrix}=\mathbf{\bar{M}}\cdot
  \begin{bmatrix}
           E_x \\
           E_y \\
           H_x \\
           H_y 
  \end{bmatrix}
\end{equation}
with
\begin{equation}
\mathbf{\bar{M}}=\begin{bmatrix}
           0&0&-\frac{k_xk_y}{\omega\epsilon_0\epsilon_{r,2}}& -\omega\mu_0+\frac{k_x^2}{\omega\epsilon_0\epsilon_{r,2}}\\
           0&0& \omega\mu_0-\frac{k_y^2}{\omega\epsilon_0\epsilon_{r,2}} &\frac{k_xk_y}{\omega\epsilon_0\epsilon_{r,2}} \\
           \frac{k_xk_y}{\omega\mu_0}&\omega\epsilon_0\epsilon_{r,2} -\frac{k_x^2}{\omega\mu_0}&0 &0  \\
          - {\omega\epsilon_0\epsilon_{r,2}} +\frac{k_y^2}{\omega\mu_0}& -\frac{k_xk_y}{\omega\mu_0} &0 &0
  \end{bmatrix}.
\end{equation}
Then, the problem of propagation in the dielectric material reduces to $i\partial_z\mathbf{f}=\mathbf{\bar{M}}\cdot\mathbf{f}$  that has a formal solution given by:
\begin{equation}
\mathbf{f}(z)=\exp(-iz\mathbf{\bar{M}})\cdot\mathbf{f}(z=0).\label{S19}
\end{equation}
The matrix $\exp(-iz\mathbf{\bar{M}})$ is a 4x4 transmission matrix as it relates the values of the fields in two dielectric material planes. Using Eqs. (\ref{S13}) and (\ref{S19}) one can write that:
\begin{equation}
\mathbf{f}(z=d^+)=\mathbf{\bar{U}}_L\cdot\mathbf{f}(z=0^+),
\end{equation}
where $\mathbf{\bar{U}}_L$ is the “one-layer” transmission matrix given by
\begin{equation}
\mathbf{\bar{U}}_L=\mathbf{\bar{U}}_B\cdot\exp(-i\mathbf{\bar{M}}d).
\end{equation}
Finally, the global transmission matrix that relates the transverse fields at the input and output interfaces,
\begin{equation}
\mathbf{f}(z=(N-1)d^+)=\mathbf{\bar{U}}_{\rm{global}}\cdot\mathbf{f}(z=0^-). \label{S22}
\end{equation}
is given by
\begin{equation}
\mathbf{\bar{U}}_{\rm{global}}=(\mathbf{\bar{U}}_L)^{N-1}\cdot\mathbf{\bar{U}}_B.
\end{equation}
The fields at the interface $z=0^-$ are given by:
\begin{equation}
\mathbf{f}(z=0^-)=\mathbf{f}^{\rm{inc}}+\mathbf{f}^{\rm{ref}}=\begin{bmatrix}
           \mathbf{E}^{\rm{inc}}_{\rm T}+\mathbf{E}^{\rm{ref}}_{\rm T}\\
           \mathbf{H}^{\rm{inc}}_{\rm T}+\mathbf{H}^{\rm{ref}}_{\rm T}
  \end{bmatrix}=
  \begin{bmatrix}
          (\mathbf{1}+\boldsymbol{\bar{\rho}}_\text{ML})\cdot \mathbf{E}^{\rm{inc}}_{\rm T}\\
         - \mathbf{\bar{J}}\cdot\mathbf{\bar{Y}}_1\cdot(\mathbf{1}-\boldsymbol{\bar{\rho}}_\text{ML})\cdot \mathbf{E}^{\rm{inc}}_{\rm T}
         \end{bmatrix}e^{i(k_xx+k_yy)}
\end{equation}
where $\boldsymbol{\bar{\rho}}_\text{ML}$  represents the reflection matrix for the multilayer system. On the other hand, the transverse fields at the interface $z=(N-1)d^+$  are given by:
\begin{equation}
\mathbf{f}(z=(N-1)d^+)=
 \begin{bmatrix}
          \mathbf{\bar{t}}_\text{ML}\cdot \mathbf{E}^{\rm{inc}}_{\rm T}\\
         - \mathbf{\bar{J}}\cdot\mathbf{\bar{Y}}_1\cdot\mathbf{\bar{t}}_\text{ML}\cdot \mathbf{E}^{\rm{inc}}_{\rm T}
         \end{bmatrix}e^{i(k_xx+k_yy)}
\end{equation}
where $\mathbf{\bar{t}}_\text{ML}$ represents the transmission matrix for the multilayer system. Therefore, from Eq. (\ref{S22}) it follows that:
\begin{equation}
\begin{bmatrix}
          \mathbf{\bar{t}}_\text{ML}\cdot \mathbf{E}^{\rm{inc}}_{\rm T}\\
         - \mathbf{\bar{J}}\cdot\mathbf{\bar{Y}}_1\cdot\mathbf{\bar{t}}_\text{ML}\cdot \mathbf{E}^{\rm{inc}}_{\rm T}
         \end{bmatrix}
=\mathbf{\bar{U}}_{\rm{global}}\cdot
 \begin{bmatrix}
          (\mathbf{1}+\boldsymbol{\bar{\rho}}_\text{ML})\cdot \mathbf{E}^{\rm{inc}}_{\rm T}\\
         - \mathbf{\bar{J}}\cdot\mathbf{\bar{Y}}_1\cdot(\mathbf{1}-\boldsymbol{\bar{\rho}}_\text{ML})\cdot \mathbf{E}^{\rm{inc}}_{\rm T}
         \end{bmatrix}.
\end{equation}
By writing the global transmission matrix as
\begin{equation}
\mathbf{\bar{U}}_{\rm{global}}=\begin{bmatrix}
\mathbf{\bar{A}}&\mathbf{\bar{B}}\\
\mathbf{\bar{C}}&\mathbf{\bar{D}}
\end{bmatrix}.
\end{equation}
where $\mathbf{\bar{A}}, \mathbf{\bar{B}},\mathbf{\bar{C}}, \mathbf{\bar{D}}$ are $2 \times 2$ matrices, it is found after straightforward algebra that the reflection and transmission matrices are given by
\begin{equation}
\boldsymbol{\bar{\rho}}_\text{ML}=(\mathbf{\bar{Q}}_1+\mathbf{\bar{Q}}_2)^{-1}\cdot(\mathbf{\bar{Q}}_1-\mathbf{\bar{Q}}_2), \quad    \mathbf{\bar{t}}_\text{ML}=2\left(\mathbf{\bar{A}}\cdot(\mathbf{\bar{Q}}_1+\mathbf{\bar{Q}}_2)^{-1}\cdot\mathbf{\bar{Q}}_1-\mathbf{\bar{B}}\cdot\mathbf{\bar{J}}\cdot \mathbf{\bar{Y}}_1\cdot(\mathbf{\bar{Q}}_1+\mathbf{\bar{Q}}_2)^{-1}\cdot\mathbf{\bar{Q}}_2\right),\end{equation}
where
\begin{equation}
\mathbf{\bar{Q}}_1= (\mathbf{\bar{J}}\cdot\mathbf{\bar{Y}}_1 \cdot \mathbf{\bar{B}}+ \mathbf{\bar{D}})\cdot \mathbf{\bar{J}}\cdot\mathbf{\bar{Y}}_1,  \quad  
\mathbf{\bar{Q}}_2=\mathbf{\bar{J}}\cdot\mathbf{\bar{Y}}_1\cdot\mathbf{\bar{A}}+ \mathbf{\bar{C}}.
\end{equation}

\section{E. Link with the response of a MOSFET transistor}

In what follows, we highlight some similarities between the material response derived in the main text and the response of MOSFET transistors.

We consider an n-channel (depletion mode) MOSFET transistor operated in the linear region \cite{Sze, Pozar}.
The drain-source channel is directed along the $y$-direction. The voltage on the gate terminal controls the electric field along the $x$-direction. The gate-to-source and the drain-to-source voltages are $V_{GS}$ and $V_{DS}$, respectively, whereas the gate and drain currents are  $I_G$ and $I_D$, respectively. It is possible to establish the following correspondence between the parameters of the 2D material (with the electric bias along $+y$) of the main text and the parameters of a MOSFET transistor:  $V_{GS} \to E_x$, $V_{DS} \to  E_y$, $I_G \to j_x$ and $I_D \to j_y$.

The response of the transistor is of the type \cite{Sze}:
\begin{eqnarray*}
 I_G&=C_{GS} \dfrac{dV_{GS}}{dt} 
 \\
 I_D&= I_D(V_{GS},V_{DS})
\end{eqnarray*}
where $C_{GS}$ is the gate-to-source capacitance and $I_D(V_{GS},V_{DS})$ is some nonlinear function of the voltages that determines the transfer function of the transistor. As is well-known, under a static bias the transistor response can be linearized. The response for a weak dynamical excitation with frequency $\omega$ is given by: 
\begin{equation}
\left( {\begin{array}{*{20}{c}}
{I_G^\omega }\\
{I_{D}^\omega }
\end{array}} \right) = \underbrace {\left( {\begin{array}{*{20}{c}}
{ - i\omega {C_{GS}}}&0\\
{{g_m}}&{{g_{22}}}
\end{array}} \right)}_{\overline {\bf{Y}} }\left( {\begin{array}{*{20}{c}}
{V_{GS}^\omega }\\
{V_{DS}^\omega }
\end{array}} \right) \label{SYmatrix}
\end{equation}
where ${g_{22}} = \frac{{\partial {I_{D}}}}{{\partial {V_{DS}}}}$ and ${g_m} = \frac{{\partial {I_{D}}}}{{\partial {V_{GS}}}}$ are determined uniquely by the bias point. The parameter $g_m$ gives the transconductance gain. The admittance matrix ${\bf{Y}}$ is the circuit counterpart of the conductivity matrix $\overline {\bf{\sigma }}$ of the main text, and Eq. (\ref{SYmatrix}) is the circuit counterpart of the formula ${\bf{j}} = \overline {\bf{\sigma }}  \cdot {\bf{E}}$. Both matrices do not have the transpose symmetry. The structure of the two matrices is slightly different (most noticeably $Y_{12}=0$ whereas $\sigma_{12}\ne0$) because the transistor is an insulator along the direction of the gate terminal ($x$-direction) due to the oxide layer. In contrast, the 2D material allows the current to flow along both the $x$ and $y$ directions. Similar to the 2D material, the transconductance gain $g_m$ is inversely proportional to the scattering rate (it is proportional to the electron mobility) \cite{Sze}. 


\section{F. Strained twisted bilayer graphene}
The band structure of strained twisted bilayer graphene can be calculated using the procedure described in the supplementary material of Ref. \cite{Pantaleon2021}. For a single graphene layer, the lattice vectors can be written as $\mathbf{a}_{1}=a(1,0)$ and $\mathbf{a}_{2}=a(1/2,\sqrt{3}/2)$ and the reciprocal lattice vectors are  $\mathbf{b}_{1}=\frac{2\pi}{a}(1,-\sqrt{3}/3)$ and $\mathbf{b}_{2}=\frac{2\pi}{a}(0,2\sqrt{3}/3)$, with $a\approx2.46 \text{\AA}$. The graphene Dirac points are $\boldsymbol{K}_{\pm}=-\xi(2\mathbf{b}_{1}+\mathbf{b}_{2})/3$ where $\xi=\pm1$ is the valley index.  At low energy, the monolayer can be described by a continuum model in the vicinity of the Dirac points:  $H(\boldsymbol{q})=-(\hbar v_{f}/a)\boldsymbol{q}\cdot(\xi\sigma_{x},\sigma_{y})$ where $\sigma_i$ are the Pauli matrices.

A twisted bilayer graphene can be obtained by rotating the two layers around a common $B$ site \citep{Moon2013}. The lower layer $l=d$ is rotated by an angle $\theta/2$ and the upper layer $l=u$ is rotated by $-\theta/2$.
The primitive and reciprocal lattice vectors in the rotated layers are written as $\mathbf{a}_{i}^{l}=\bar{\mathbf{R}}(\mp\theta/2)\mathbf{a}_{i}$ and $\mathbf{b}_{i}^{l}=\bar{\mathbf{R}}(\mp\theta/2)\mathbf{b}_{i}$ where $\bar{\mathbf{R}}(\theta)$ is a rotation matrix by an angle $\theta$.  
The twisted bilayer graphene presents strain that originates from the interaction with the $h$-BN substrate and top layer. 
If we consider uniaxial deformations, the system is  stressed along one direction and unstressed on the perpendicular direction \citep{Bi2019}, breaking the $C_3$ symmetry. To include uniaxial strain perpendicular to the zigzag direction, the linear strain tensor $\bar{\boldsymbol{\varepsilon}}$ with  relative magnitude $\epsilon$ can be written as 
\begin{equation}
\bar{\boldsymbol{\varepsilon}}=\epsilon\left(\begin{array}{cc}
-1 & 0\\
0 & \nu_0
\end{array}\right),\label{strain}
\end{equation}
with $\nu_0=0.16$ being the Poisson ratio for graphene. In our case $\bar{\boldsymbol{\varepsilon}}^{T}=\bar{\boldsymbol{\varepsilon}}$ and, the transformed primitive and reciprocal lattice vectors for each rotated graphene layer are given by $\boldsymbol{\alpha}_{i}^{l} =(\mathbb{1}+\bar{\boldsymbol{\varepsilon}}_{l})a_{i}^{l}$ and $\boldsymbol{\beta}_{i}^{l}  =(\mathbb{1}-\bar{\boldsymbol{\varepsilon}}_{l})\mathbf{b}_{i}^{l}$,
where $\bar{\boldsymbol{\varepsilon}}_{l}$ is the strain tensor and $l$ a layer index. In TBG with uniaxial heterostrain, the
strain is applied in opposite directions in the two graphene layers. The relative deformation is given by $\bar{\boldsymbol{\varepsilon}}=\bar{\boldsymbol{\varepsilon}}_{u}-\bar{\boldsymbol{\varepsilon}}_{d}$ with $\bar{\boldsymbol{\varepsilon}}_{u} = -\bar{\boldsymbol{\varepsilon}}_{d} =\frac{1}{2}\bar{\boldsymbol{\varepsilon}}$.  The strain modifies the intralayer Hamiltonian and consequently changes the electronic structure. For small strain, one can introduce a pseudomagnetic field in the low energy Hamiltonian \citep{Nam2017}. In each graphene layer, the vector potential, $\mathbf{A}^{l}=(A_{x}^{l},A_{y}^{l})$, for this field is given by  $A_{x}^{l} =\frac{\sqrt{3}}{2a}\beta_{G}\left(\varepsilon_{xx}^{l}-\varepsilon_{yy}^{l}\right)$  \citep{Guinea2009},
where $\beta_{G}\approx3.14$ is a dimensionless parameter. In small angle TBG,  the moir\'e superlattice constant is much larger than the atomic scale.  The low energy physics is still restricted to  the vicinity of the Dirac points. This allows us to study each valley separately with the continuum model of Bistritzer and MacDonald \cite{Bistritzer2011}.  However, the strain shifts the Dirac points from their original positions to  $(\mathbb{1}-\bar{\boldsymbol{\varepsilon}})\boldsymbol{K}_{\xi}-\xi A^{l}$. In the layer basis, the effective Hamiltonian for the valley $\xi=\pm1$ has two Dirac Hamiltonian terms for each graphene layer and a tunneling term describing the hopping between them. It can be written as
\begin{equation}
H_\xi=\left(\begin{array}{cc}
H(\boldsymbol{q}_{d,\xi})+\Delta_{d}\sigma_{z} & U^{\dagger}\\
U & H(\boldsymbol{q}_{u,\xi})+\Delta_{u}\sigma_{z}
\end{array}\right),\label{Htbg}
\end{equation}
where $\boldsymbol{q}_{l,xi}=\bar{\mathbf{R}}(\pm\theta/2)(\mathbb{1}+\bar{\boldsymbol{\varepsilon}}_{l})(\boldsymbol{q}-\mathbf{D}_{l,\xi})$
with $\pm$ for $l=d$ and $l=u$, respectively.  $H(\boldsymbol{q})$ is the Hamiltonian for a monolayer graphene. $\Delta_{d/u}$ is the staggered potential induced by the $h$-BN substrate and top layer \citep{Cea2020}, which is responsible for breaking inversion symmetry and opening a band gap in monolayer graphene. Finally, the off-diagonal term $U$ describes the interlayer coupling between graphene layers. In the low energy approximation, it is given by the Fourier expansion:
\begin{align}
U & =\left(\begin{array}{cc}
u & u^{\prime}\\
u^{\prime} & u
\end{array}\right)+\left(\begin{array}{cc}
u & u^{\prime}\omega^{-\zeta}\\
u^{\prime}\omega^{\zeta} & u
\end{array}\right)e^{i\zeta\boldsymbol{g}_{1}\cdot\boldsymbol{r}} 
  +\left(\begin{array}{cc}
u & u^{\prime}\omega^{\zeta}\\
u^{\prime}\omega^{-\zeta} & u
\end{array}\right)e^{i\zeta(\boldsymbol{g}_{1}+\boldsymbol{g}_{2})\cdot\boldsymbol{r}}
\label{Ut} 
\end{align}
where $\omega=e^{2\pi i/3}$, with $u=0.0797$ eV and $u^{\prime}=0.0975$ eV \citep{Koshino2018}. The large periodicity of the moir\'e superlattice leads to a small periodicity in the reciprocal space represented by the moir\'e Brillouin zone (mBZ).  $\boldsymbol{k}$ is then restricted to the first mBZ. We solve the secular equation $H(\boldsymbol{k})\left|\Psi_{n\boldsymbol{k}}\right\rangle =E_{n}(\boldsymbol{k})\left|\Psi_{n\boldsymbol{k}}\right\rangle $ to obtain the  eigenvectors and energies $\left|\Psi_{n\boldsymbol{k}}\right\rangle $ and $E_{n}(\boldsymbol{k})$. $\boldsymbol{q}=\boldsymbol{k}+n\boldsymbol{g}_{1}+m\boldsymbol{g}_{2}$,
where $m,n$ are integers.
\begin{figure}[h!]
	\centering
	 \includegraphics[width=0.6\linewidth,clip]{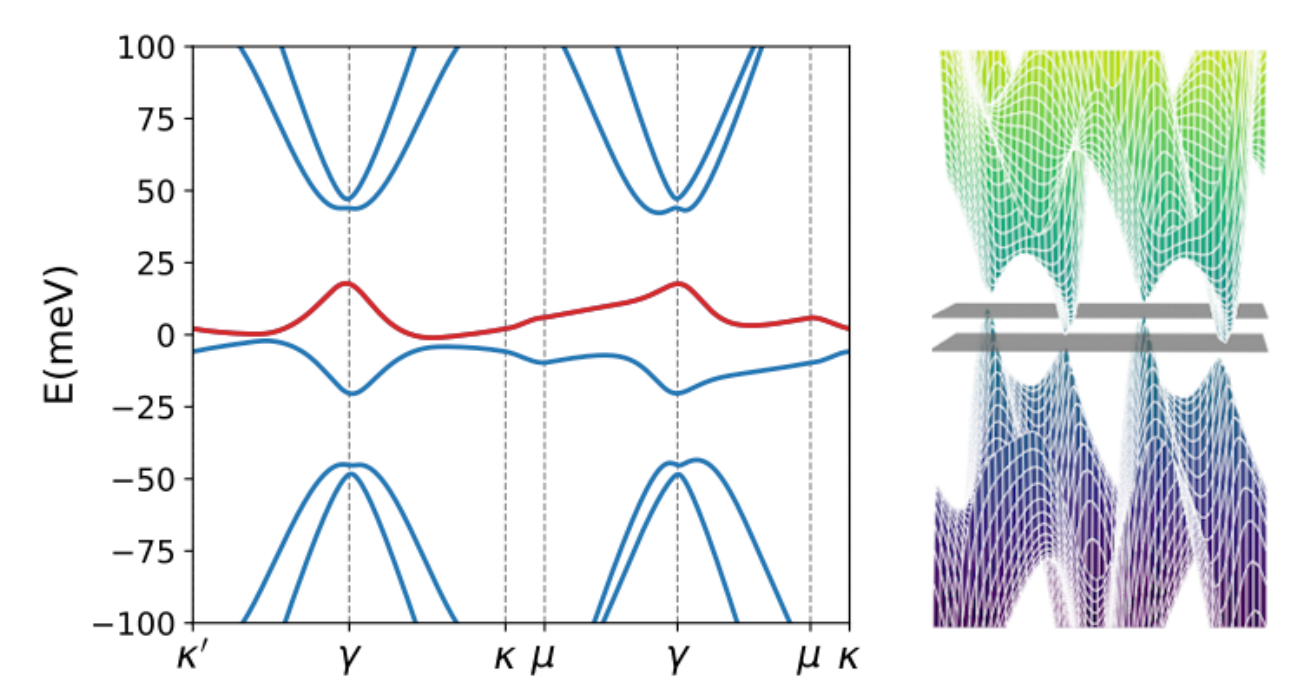}   
	\caption{Left: Band structure of a strained twisted bilayer graphene for $\theta=1.25^{\circ}$, uniaxial strain $\epsilon=0.3\%$ and   $\Delta_d=-\Delta_u=1 $ meV. $\boldsymbol{k}$ is restricted to the first mBZ. The band that originates the main contribution to the Berry curvature dipole for the range of Fermi  energies used in the main text is highlighted in red. Right: Three dimensional view of the band edges for the same band structure as in the left panel. The gray planes give the locations of the valence and conduction band edges.} 
	\label{figS4}
\end{figure} 
The energy dispersion and eigenvectors are obtained by diagonalizing the Hamiltonian of equation (\ref{Htbg}) numerically. The number of Fourier components that define the eigenvectors in the numerical calculation is chosen as the minimal number to obtain convergence for the low energy moir\'e bands.  

We calculate the energies and wave-functions for a TBG with a heterostrain of 0.3 $\%$, a twist angle of $1.25^\circ$ degrees and opposite staggered potentials $\Delta_d=-\Delta_u$=1 meV for each graphene layer. Following the procedure outlined in the previous paragraphs, we obtain the moir\'e band structure of Fig. \ref{figS4}. One can see that the staggered potential leads to energy band anti-crossings with band separations of the order of few meV. This results in large Berry curvatures that could also occur in other graphene systems where a large periodicity is engineered. The right panel of Fig. \ref{figS4} shows the position of the band edges in the mBZ, which are  not located in the path connecting the high symmetry points. The conduction band edge is lower than the valence band edge, closing the gap. 

The band structure and the eigenstates are used to calculate the Berry curvature dipole and the optical conductivity of a TBG without any further approximation. The Berry curvature is calculated numerically using Fukui's method \cite{Fukui2005}. For multiband systems, the components of the BD are obtained using
\begin{equation}
D_{a}=\sum_n\int\frac{d^2k}{(2\pi)^2}\Omega^z_n\frac{\partial f^0_\mathbf{k}}{\partial k_a}.
\end{equation}
where $n$ is the band index. The contributions of the two valleys are summed. 
The DC conductivity is given by ($v_\mathbf{k}^\alpha =\pd{\ek}{k_\alpha}$)
\begin{equation}
\sigma_{\alpha\beta}=-\frac{e^2}{\gamma}\int\frac{d^2k}{(2\pi)^2}v_\mathbf{k}^\alpha v_\mathbf{k}^\beta \left(\pd{f^0_\mathbf{k}}{\ek}\right),
\end{equation}
and is obtained with numerical integration. For Fermi energies lying at the bottom of the band  highlighted in red in Figure \ref{figS4}, the direct gap to the upper band is larger than 100 meV. Thus, for the frequency ranges considered in the main text, interband contributions to the optical conductivity are negligible. 

\begin{figure}[h!]
	\centering
	 \includegraphics[width=0.6\linewidth,clip]{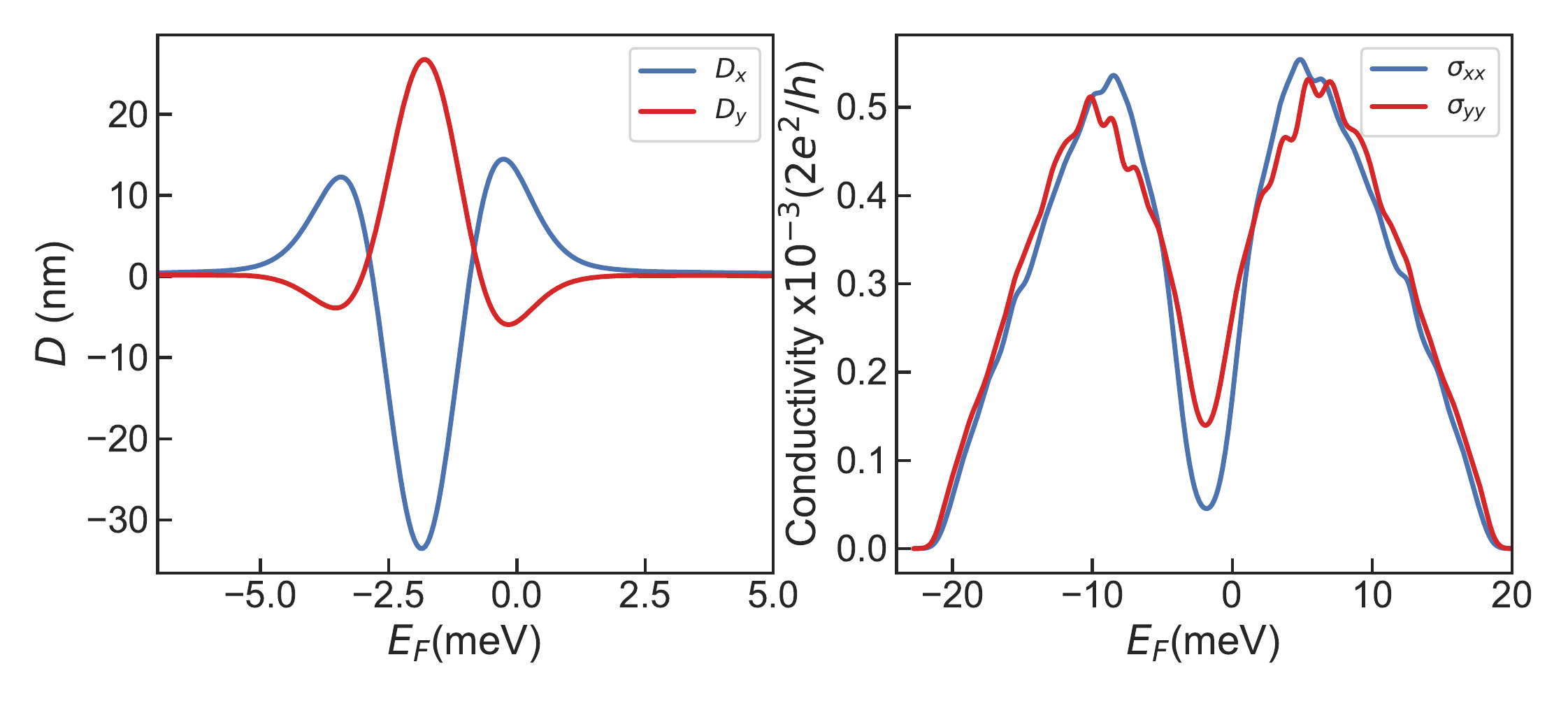}   
	\caption{Berry curvature dipole (left) and longitudinal DC conductivity (right) as a function of the Fermi energy for a strained twisted bilayer graphene for $\theta=1.25^{\circ}$ and uniaxial strain $\epsilon=0.3\%$.  $\Delta_d=-\Delta_u= 1$ meV, $\gamma$=2 THz and $T$=10K.} 
	\label{figS5}
\end{figure} 

The BD and DC conductivity for the same parameters as in Fig. \ref{figS4} are given in Fig. \ref{figS5} as a function of the Fermi energy. Figure \ref{figS5} shows the two Cartesian components of the  BD. The BD is maximized for a energy window where the Fermi surface contains electron and hole pockets and both contribute to the BD. In two dimensions, the Berry curvature dipole behaves as a pseudo-vector $\mathbf{D}$ in the $xy$ plane. In the main text, the electric field direction and the $y$-axis orientation were chosen to be parallel to $\mathbf{D}$.


\end{document}